\documentclass[aps,twocolumn]{revtex4}
\usepackage{graphicx,subfigure}
\usepackage{amsmath,amssymb}
\usepackage{bm}

\newcommand{\be}{\begin{eqnarray}}
\newcommand{\ee}{\end{eqnarray}}

\def\slashchar#1{\setbox0=\hbox{$#1$}           % set a box for #1 
   \dimen0=\wd0                                 % and get its size
   \setbox1=\hbox{/} \dimen1=\wd1               % get size of /
  \ifdim\dimen0>\dimen1                        % #1 is bigger
 \rlap{\hbox to \dimen0{\hfil/\hfil}}      % so center / in box
  #1                                        % and print #1
 \else                                        % / is bigger
    \rlap{\hbox to \dimen1{\hfil$#1$\hfil}}   % so center #1
    /                                         % and print /
 \fi}                                         %

\begin{document}

\title{ Light-front wave functions of mesons, baryons and pentaquarks,\\
with topology-induced local 4-quark interaction}

\author{ 
Edward  Shuryak }

\affiliation{Department of Physics and Astronomy, Stony Brook University,
Stony Brook NY 11794-3800, USA}

\begin{abstract}
We calculate light-front wave functions of mesons, baryons and pentaquarks
in a model including constituent mass (representing chiral symmetry breaking),  harmonic confining potential,  and  4-quark local interaction of 't Hooft type. The model
is a simplified version of that used by Jia and Vary.
The method used is numerical diagonalization of the Hamiltonian
matrix, with certain functional basis. We found that the nucleon wave function displays strong diquar correlations, unlike that for Delta (decuplet)
baryon.  We also calculate 3-quark-5-quark admixture to baryons,
and the resulting antiquark sea PDF.
\end{abstract}
\maketitle

\section{Introduction}
\subsection{Various roads towards hadronic properties}
%Studies of mesons and baryons proceed in several different directions. 
Let us start with a general picture, describing various approaches in 
the theory of hadrons, identifying the following:\\

(i){\em  Traditional quark models} (too many to mention here) are aimed at calculation of 
static properties (e.g. masses, radii, magnetic moments etc). Normally all calculations are done
in hadron's rest frame, using certain model Hamiltonians.  Typically, chiral symmetry breaking is included via
effective ``constituent quark" masses, the Coulomb-like and confinement  forces are included via some 
potentials. In some models also  a ``residual" interaction is also included, via some 4-quark terms of the NJL type.\\

(ii) Numerical calculation of  {\em Euclidean-time} two- and three-point  correlation functions is another general approach, 
with a source and sink operators creating a state with needed quantum numbers, and the
third one in between, representing the observable. Originally started from small-distance
OPE and the QCD sum rule method, it moved to intermediate distances (see review
e.g. in  \cite{Shuryak:1993kg}), and is now mostly used at large time separations $\Delta \tau$ 
(compared to inverse mass gaps in the problem $1/\Delta M$) by the lattice gauge theory (LGT) simulations.  This condition ensure ``relaxation" of the correlators 
to the lowest mass hadron in a given channel.\\

(iii) Light-front quantization using also certain model Hamiltonians, aimed at
 the set of quantities, available from experiment. 
Deep inelastic scattering (DIS), as well as many other hard processes,
use factorization theorems of perturbative QCD and
 nonperturbative {\em parton distribution functions} (PDFs). Hard exclusive processes
(such as e.g. formfactors) 
are described in terms of  nonperturbative  hadron on-light-front wave functions (LCWFs), for reviews see \cite{Brodsky:1989pv,Chernyak:1983ej}.

(iv) Relatively new approach is ``holographic QCD", describing hadrons as quantum fields 
propagating in the ``bulk" space with extra dimensions. It originally supposed to 
be a dual description to some strong-coupling regime of QCD, and therefore was mostly
used for description of Quark-Gluon Plasma (QGP) phase at high temperatures. 
Nevertheless, its versions including confinement (via dilaton background with certain ``walls"
\cite{Erlich:2005qh} ) and quark-related fields (especially in the so called Veneziano limit in which both the number of flavors and colors are large $N_f,N_c\rightarrow \infty, N_f/N_c=fixed$ \cite{Jarvinen:2011qe} )
  do reproduce hadronic spectroscopy, with nice Regge trajectories. 
  The holographic models also led to interesting revival of baryons-as-solitons type models, generalizing skyrmions and including also vector meson clouds.  
  Brodsky and de Teramond  \cite{Brodsky:2006uqa} proposed to relate the wave functions in extra dimension $z$ to those on the light cone,
identifying  $z$ with certain combination of the light cone variables $\zeta$. 
 Needless to say, all of these are  models constructed ``bottom-up", but with well defined Lagrangians and some economic set of parameters, from
 which a lot of (mutually consistent) predictions can be worked out.

While (i) and (iv) remain basically in realm of model building,
  (ii) remains the most fundamental and
consistent approach. Lattice studies, starting  from the first principles
of QCD had convincingly demonstrated that they correctly include all nonperturbative phenomena.  They do display chiral symmetry breaking and confinement, and reproduce
accurately
hadronic masses. Yet its contact with PDFs and light-front wave functions remains difficult. The light front direction, on the other hand, for decades relied on perturbative QCD,
in denial of most of nonperturbative physics. 

The aim of the present paper is to bridge the gap between approaches (ii) and (iii).  It is
 a kind of pilot project, using a particular quark model 
(deliberately stripped to its ``bare bones") and then performing consistent dynamical 
calculations of the light front wave functions in it. It follows directly  
the approach of Jia and Vary Ref\cite{Jia:2018ary} to pion and rho mesons. 
We extended its applications to baryons, Delta (decuplet) and the proton (the octet),
the 5-quark systems (pentaquarks), and finally to 3 quark - 5 quark mixing,
dynamically addressing the issue of the non-perturbative antiquark sea.

The model Hamiltonian has three terms: (i) the constituent mass term, representing the chiral symmetry breaking; (ii) harmonic-type confining potential; and last but not least (iii)
  the 4-quark local interaction of
Nambu-Jona-Lasino (NJL) type. One simplification we use is to consider only the 
longitudinal degrees of freedom, ignoring the transverse motion. Another is to
reduce the complicated NJL operator to a single topology-induced 't Hooft vertex \cite{tHooft:1976snw}
. This latter
step is explained in the next subsection. 

\subsection{Topology-induced multiquark interactions}

Nambu and Jona-Lasinio 1961 paper \cite{Nambu:1961tp} was an amazing breakthrough. Before the word ``quark" was invented, and
one learned anything about quark masses, it postulated the notion of chiral symmetry and its
spontaneously breaking. They postulated existence of 4-fermion interaction, with some coupling $G$, strong enough to make a superconductor-like gap even in fermionic vacuum. The second important
parameter of the model was the cutoff $\Lambda \sim 1\, GeV$, below which their hypothetical
attractive  4-fermion interaction operates. 
%For recent application of NJL model see e.g. \cite{wise}.

\begin{figure}[h!]
\begin{center}
\includegraphics[width=8cm]{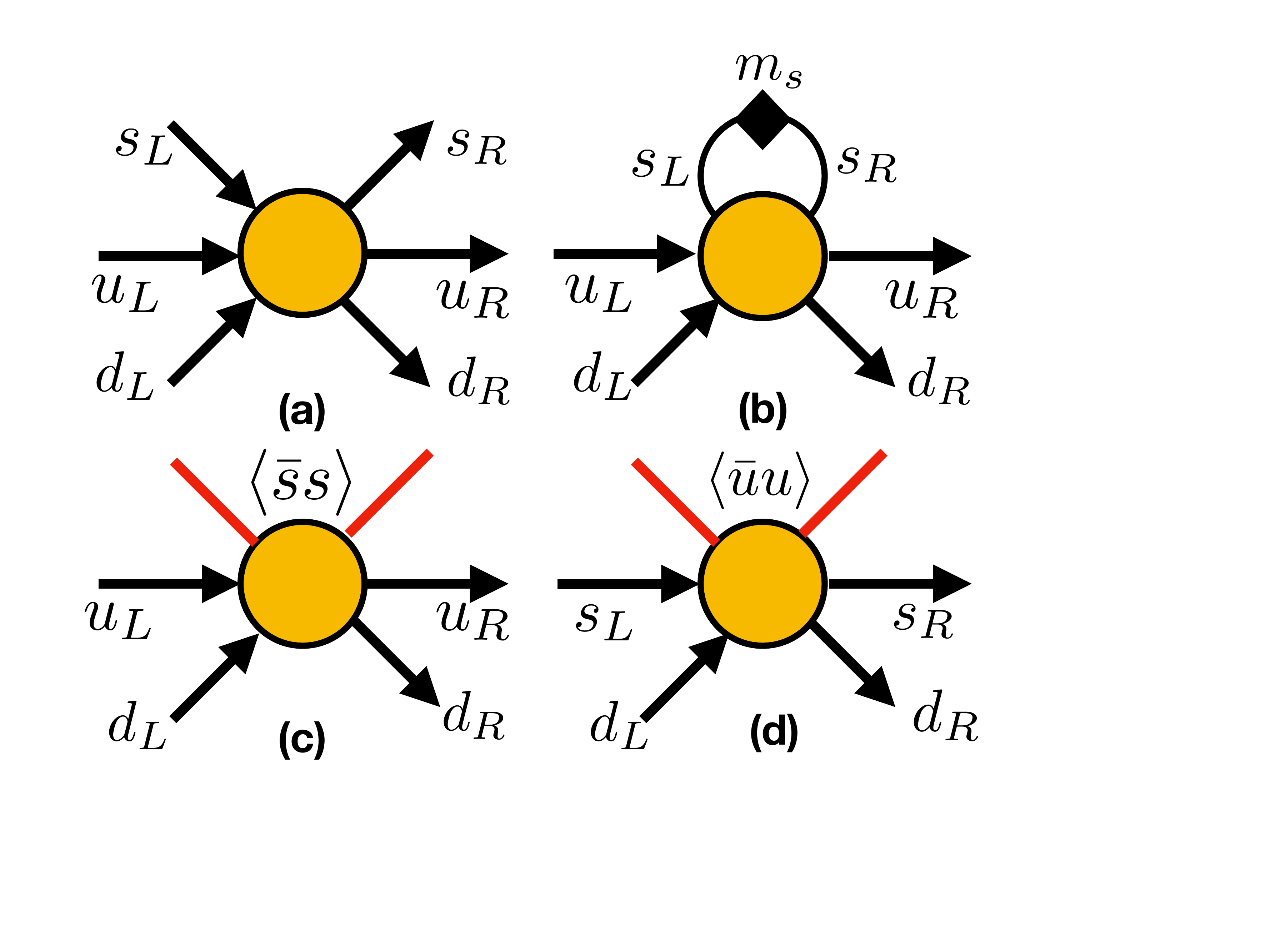}
\caption{Schematic form of the 6-quark 't Hooft effective Lagrangian is shown in fig (a). If quarks are
massive, one can make a loop shown in (b), reducing it to 4-fermion operator. Note a black rhomb indicating  the mass insertion into a propagator. We only show it for $s$ quark, hinting that for $u,d$ their masses are too small to make such
diagram really relevant. In (c,d) we show other types of effective 4-fermion vertices, appearing because 
some quark pairs can be absorbed by a
nonzero quark condensates (red lines). }
\label{fig_multiquark}
\end{center}
\end{figure}

After discovery of QCD, gauge field monopoles and instantons, a very curious relations was found \cite{tHooft:1976snw}, between the Dirac operator and background
gauge topology: they have certain zero modes related to the topological charge. This mathematical phenomenon has 
direct physical consequences, multi-quark interaction vertex described by the so called 't Hooft effective    
Lagrangian. Since in QCD it includes all three flavors of light quarks, $u,d,s$, it is a 6-quark effective vertex, 
schematically shown in Fig. \ref{fig_multiquark}(a). Note its key feature,  opposite chiralities $L,R$ of the incoming
and outgoing quarks: it is so because in order to have zero modes of the Dirac equation, quarks and antiquarks should have the same magnetic moments.  Unlike vectorial interaction with non-topological
glue, this Lagrangian directly connect left and right components of quark fields, explicitly breaking $U(1)_a$ 
chiral symmetry.

With the advent of the instanton liquid model  (ILM) \cite{Shuryak:1981ff} it became clear that it provides an explanation
to the origins of
hypothetical NJL interaction. The NJL coupling $G_{NJL}$ and cutoff $\Lambda_{NJL}$
were substituted by other two parameters, the instanton
4d density $n$ and the typical size $\rho$. Like the $NJL$ action, t' Hooft effective action also preserves  the $SU(N_f)$ chiral symmetry, but  is also strong enough to break it spontaneously,
creating nonzero quark condensates $\langle \bar s s \rangle, \langle \bar u u \rangle\neq 0$
appearing in diagrams (c,d) of Fig.\ref{fig_multiquark}. The residual 4-quark $(\bar u u )(\bar d d)$ interaction,
induced by the diagrams (b,c), is the one to be used below.
Note, that
unlike the NJL action, it explicitly breaks the chiral $U_c(1)$ symmetry.
  
In 1990's the so called 
``interacting instanton liquid model", which numerically solved 
the vacuum properties using 't Hooft Lagrangian to all orders, providing
hadron spectroscopy and Euclidean correlation functions, for review see \cite{Schafer:1996wv}. Recent advances to  finite temperatures and
 QCD phase transitions at finite temperature is based on instanton constituents, the ``instanton dyons: we do not go into that here and only comment that the structiure of the  't Hooft Lagrangian
 remains the same. 

In this pilot study we would simplify this residual 4-fermion interaction, as much as possible,
 assuming that 't Hooft effective action is $local$. This implies that
the instanton radius (or $1/\Lambda_{NJL}$) is much smaller than typical hadronic sizes,
$\rho\ll R_{hadrons}$ and therefore can be neglected. With this assumption, the residual
4-fermion interaction has only one parameter, the coupling.    

It needs to be stated that such simplification goes with a certain price: the wave functions get singular at $x\rightarrow 1$ causing bad convergence of their expansion in
basis functions: we will terminate series  by hand. Let us add, that small size of 
QCD topological objects also produce technical difficulties for lattice simulations:
many quantities (and PDF moments are among them, see e.g. \cite{Bali:2019ecy}) show significant dependence on the value of lattice spacing $a$ down to very fine lattices,
with $a\sim 1/(2-3\, GeV)$, so coninuum extrapolation $a\rightarrow 0$ is a nontrivial step.

Small sizes of instanton and instanton-dyons explain few other puzzles, known in
hadronic physics and by
lattice practitioners. We will not discuss phenomena related to strange quark mass 
in this work, but notice in passing, the
 ``puzzle of strong breaking of the  $SU(3)$ flavor symmetry".  
 Naively, in NJL-like models $$m_s\sim 0.1\, GeV\ll \Lambda_{NJL}\sim 1\, GeV$$ is a small parameter, and expansion in it should
 be well behaved. However, it is far from being seen in the real and numerical data.
 One particular 
manifestation of it, observed e.g. in recent lattice work already mentioned \cite{Bali:2019ecy}, 
is that PDF moments for various
octet baryons  $N,\Lambda,\Sigma$ are very different. The $SU(3)$ flavor symmetry is broken at 100\% level.  
%This is rather surprising, taking into account that
%the ratio of the strange quark mass to the relevant momentum scale  is small $m_s/
%\Lambda_{NJL} \sim 1/10$.
Its explanation in ILM is also related to small sizes of the topological objects. The 4-fermion vertex of 
Fig\ref{fig_multiquark}(b) is proportional to (chirality-flipping) factor $m_s\rho$. 
(appearence of $\rho$ is obvious from 
dimensional consideration.)
The effective vertices generated by diagrams
(c,d) are proportional to the respective quark condensates, as indicated. For convenience,
the sum of (b) and say (c) vertex is written in a form $(m_s+M_s^*)\rho$ where the so called 
{\em determinantal mass} $M^*$ 
(not to be confused with the constituent quark mass)
can be calculated for small-size instanton a la OPE to be
\be M_f^*= {2\pi^2 \over 3} | \langle \bar q_f q_f \rangle | \rho^2 \ee
and, by coincidence of numbers, is comparable to $m_s$. Therefore, the effective
4-fermi interactions for light and strange quarks turns out to be factor two different! )

Although we would not discuss chiral symmetry breaking and chiral perturbation theory below, it is tempting to mention one more observation, very puzzling to
lattice practitioners. It is related not with strange quark mass, but with the light quark masses $m_u,m_d$. Those are only few $MeV$ in magnitude and, naively, 
a chiral extrapolation to $m_u,m_d\rightarrow 0$ should just be linear. Yet in practice,
 many observables show nonlinear mass dependences.
The answer, at least the one coming from ILM, is that the chiral condensate, made of collectivized instanton zero modes,
has a spread of Dirac eigenvalues proportional to $overlaps$ of zero modes of individual instantons.
To form a condensate, quark needs to hop from one instanton to the next.
The overlap of their zero modes is surprisingly small because the topology ensemble is  rather $dilute$
\be \Delta \lambda \sim   {\rho^2 / R^3} \sim 20 \, MeV\ee
comparable to quark masses used on the lattice.
Here $R\sim 1\, fm$ is a typical distance between instantons. It is $1/R^3$ because
such is the propagator of massless quark in 4d, and two $\rho$ factors
are two couplings of a quark to two instantons.

 \subsection{Comments on perturbative evolution and twist expansion}
 
 Completing the Introduction, we would like to add some comments clarifying
 relations between the model calculation of the  LFWFs, presented below and in other works, 
 with studies of the perturbative processes, involving gluon and quark-antiquark
 production. While generic by themselves, these comments had created discussion
 in literature, so perhaps they are worth to be repeated.
 
 The hadrons -- e.g. a pion or a fully polarized proton -- are of course single quantum states,
 described   by their multibody wave functions. On the other hand, the PDFs we used to deal with
 using the parton model are {\em single-body density matrices}. In principle, they
 should be calculated from the wave functions, by integrated out variables relating to
 other partons. A single struck parton, being a subsystem, posesses {\em entanglement
 entropy} with the rest of the state, as recenly pointed out by Kharzeev and Levin \cite{Kharzeev:2017qzs}.
 
 Furthermore, the celebrated and widely used DGLAP evolution equation is 
 basically a version of Boltzmann kinetic equation, with ``splitting functions" describing
 the gain and loss in PDFs as the resolution $Q^2$ changes. As for any Boltzmann equation,
 one can define the (ever increasing) entropy. Also important to note, as any other single-body  Boltzmann equation, DGLAP is based on implicit assumption that higher correlations between
 bodies are small and can be neglected. 
 Therefore, it may appear strange that DGLAP is successfully applied to the nucleon, in which
 diquark correlations are known to be so strong, that 
 it often is treated as 2-body system (e.g. the nucleon Regge trajectories
 are nearly the same as for mesons). 
 
 To understand that there is no contradiction, one needs to have a look at
 evolution including not only the {\em leading twist} operators, whose pQCD
 evolution is described by DGAP, but think of higher twist operators as well. 
 In particular, general OPE expansion for deep-inelastic scattering
  include gluon-quark and even 4-fermion operators, see e.g. \cite{Shuryak:1981kj}.
  Their matrix elements is precisely the place where inter-parton correlations are
  included. However, these operators are suppressed by powers of $1/Q^2$,
  and are normally ignored in DGLAP applications for hard processes at large $Q$. 
  
  Three lessons from this discussion: (i) A nucleon has no entropy, but its PDFs have.
  (ii) DGLAP is a kinetic equation, ignoring correlations between partons but explaining the entropy growth. It is good at high $Q$.
  (iii) Attempting to build a bridge between hadronic models and pQCD description, one thinks about the scale $Q\sim 1\, GeV^2$, where correlations and higher twist effects need to be included. Observing parton correlations is hard, not done yet, but still in principle possible.
 
 \section{Few-body kinematical variables on the light front } \label{sec_kinematics}
Textbook introduction to few-body quantum mechanics starts usually with
introduction of Jacobi coordinates, eliminating the center-of-mass motion. 
Those can be used for transverse coordinates, but not for momentum fractions $x_i,i=1..N$.
In this section we propose a set of $N-1$ coordinates
in which one can conveniently describe dynamics with longitudinal momentum fractions $x_1...x_N$, with the
 constraint
 \be \sum_{i}^N x_i=1, \,\,\,\,\,\,\,  \label{eqn_constraints} \ee 

 But before we do so, let us comment that the number of particles $N$ and variables $N-1$
 in the problems to be dealed with
 need not to be fixed.  Although such situation is unlike to what is usually described in textbooks,  it is rather common in realistic
 applications of quantum mechanics. For example,  complicated atoms or nuclei are described
 in a basis with closed shells, plus  several nucleons (or electrons) above it, plus an indefinite number particle-hole pairs. LFWS similarly should have indefinite number of quark-antiquark
 pairs.

 In general,  light front wave functions (LFWF) consist of sectors with $N$ particles, with $N$ changing from
 some minimal number $N_{min}$ (2 for mesons, 3 for baryons) to infinitity. The variables should satisfy  in each sector. 
 
 In this paper we will make some drastic simplifications, in particular:\\
(i) we will  ignore transverse momenta, as already mentioned, and focus on 
 the LFWF dependence on the longitudinal momentum fractions $x_i$ only.\\
 (ii) as we focus on 4-quark interactions, we will ignore gluons and processes involving them. So  
  our particles will be only quarks. 
  They would be dressed due to chiral symmetry breaking and thus have
  effective ``constituent quark masses". \\
  (iii) while each $N$-sector will be described by a space with certain appropriate number
  of basis states, the corresponding polynomials, we will truncate the Hilbert space to
  minimal and next-to-minimal ($N_{min}$ and $N_{min}+2$)  sectors.  While the former
  dominate the valence quark structure functions, the latter will be needed to discuss the ``sea"
  quarks and antiquarks.

  So, our
 operational Hilbert state would
including  states with 2 and 4 quarks for mesons, and 3 and 5  for baryons. Let us discuss these sectors 
subsequently, explaining   
 more convenient variables and the integration measure for each sector. In selecting the kinematical variables
 one would like to include the constraint explicitly, yet keeping the integration measure $factorizable$ and with
 the boundaries 
  
  {\bf 2-particle sector} is the simplest case.  With two momentum fractions, $x_1,x_2$ and the constraint (\ref{eqn_constraints}) there remains a single variable. For reasons to become clear soon, we select it to be
  their difference
  \be s= x_1-x_2 \ee
  in terms of which
  \be x_1={1+s \over 2}, \,\,\, x_2={1- s \over 2} \ee
  The commonly used integration measure take the form 
  \be \int dx_1 dx_2 \delta(x_1+x_2-1) x_1 x_2...= \int_{-1}^1 ds {(1-s) (1+s) \over 4}... \ee
  The polynomials used in this case, with natural weight $(1-s)^2$, are Gegenbauer 
  polynomials $C^{3/2}_n(s)$ or Jacobi $P_n^{1,1}(s)$. 
  
   {\bf 3-particle sector} has been discussed extensively in literature such as \cite{Braun:1999te}. Two kinematical variables
   suggested in this case are
   \be  s= {x_1-x_2 \over x_1+x_2}, \,\,\,\,\, t=x_1 + x_2 - x_3 \ee
   in terms of which 
   \be x_1={ (1+s) \over 2} {(1+t )\over 2}, \,\,\, x_2={ (1-s) \over 2} {(1+t) \over 2},    \ee
 and the integration measure   
     \be \int (\prod_i dx_i ) \delta(\sum_i x_i-1) (\prod_i x_i )...= \ee $$\int_{-1}^1\int_{-1}^1 ds dt {(1-s) (1+s) (1-t) (1+t)^3 \over 2^5}... $$
     is indeed factorized. Therefore one can split it into two and select appropriate  functional basis.
In its choice  we however would deviate from that in \cite{Braun:1999te} (because we do not consider perturbative renormalization and anomalous dimensions of the operators) and  will use 
\be \psi_{n,l}(s,t)\sim P_n^{1,1}(s) P_l^{1,3}(t)  \label{3part_basis} \ee

{\bf 4-particle sector} follows the previous example. Three variables are defined as 
$$ s = { x_1 - x_2 \over x_1 + x_2}, \,\,\,\, 
t = {x_1 + x_2 - x_3 \over x_1 + x_2 + x_3} , $$
\be u = x_1 + x_2 + x_3 - x_4 \ee
producing the integration measure 
 \be \int (\prod_i dx_i ) \delta(\sum_i x_i-1) (\prod_i x_i )...= \ee 
 $$\int_{-1}^1\int_{-1}^1\int_{-1}^1 ds dt du {(1 - s) (1 + s) (1 - t) (1 + t)^3 (1 - uu) (1 + uu)^5\over 32768}... $$
The functional basis is then 
\be \psi_{n,l,m}(s,t)\sim P_n^{1,1}(s) P_l^{1,3}(t)  P_m^{1,5}(u)\ee

{\bf 5-particle sector} is our last example: now the four variables, collectively called $z_i,i=1,2,3,4$ are defined by 
\be   s = { x_1 - x_2 \over x_1 + x_2}, \,\,\,\, 
t = {x_1 + x_2 - x_3 \over x_1 + x_2 + x_3} , \ee $$
 u = {x_1 + x_2 + x_3 - x_4 \over x_1 + x_2 + x_3 + x_4 } , \,\,\, w=x1 + x2 + x3 + x4 - x5 $$
 The principle idea can also be seen from the inverse relations

$$ x_1  = {1 \over 2^4} (1 + s) (1 + t) (1 + u) (1 + w) $$
$$ x_2  =  {1 \over 2^4} (1 - s) (1 + t) (1 + u) (1 + w) $$
$$ x_3  =  {1 \over 2^3}(1 - t) (1 + u) (1 + w), $$
\be x_4  =  {1 \over 2^2} (1 - u) (1 + w) \ee
$$ x_5  = 1 - x_1 - x_2 - x_3 - x_4= {1 \over 2} (1-w) $$

 The integration measure follows the previous trend, being factorizable
\be  \int {ds dt du dw \over 16777216} (1 - s) (1 + s) (1 - t) (1 + t)^3 \ee $$\times (1 - u) (1 + u)^5 (1 - 
   w) (1 + w)^7... $$
   The orthonormal polynomial basis to be used is by now obvious,
   \be  \psi_{lmnk}(s,t,u,w)\sim P_l^{1,1}(s)P_m^{1,3}(t)P_n^{1,5}(u)P_k^{1,7}(w) \label{eqn_basis} \ee
   with normalization constants determined numerically.
   
 \section{Mesons as two-quark states }  \label{sec_mesons}
 LFWFs for pion and rho mesons has been studied in Jia-Vary (JV) paper \cite{Jia:2018ary}. Their  Hamiltonian has
 four  terms including (i) the effective quark masses coming from chiral symmetry breaking; (ii) the longitudinal confinement;
 (iii) the transverse motion and confinement;   and, last but not least, (iv) the NJL 4-quark effective interaction
 \be H=H_M+H_{conf}^{||}+ H_{conf}^\perp+H_{NJL} \ee
 $$ H_M={M^2 \over x_1}+{\bar M^2 \over x_2} $$
 $$ H_{conf}^{||}={\kappa^4 \over (M+\bar M)^2} {1 \over J(x) }\partial_x J(x) \partial_x $$
 $$ H_\perp=\vec k_\perp^2 \big({1 \over x_1}+{1 \over x_2}\big) +\kappa^4 x_1x_2 \vec r_\perp^2 $$
where $M,\bar M$ are masses of quark and antiquark, $\kappa$ is the confining parameter, $J(x)=x_1x_2=(1-s)^2/4$
is the integration measure, $\vec k_\perp,\vec r_\perp$ are transverse momentum and coordinate. 
 If the masses are the same, however,
 one can simplify it $${1\over x_1}+{1 \over x_2}={1 \over x_1x_2}={4 \over (1-s^2)}  $$
 and therefore the matrix element of $H_M$ simply lack the factor $(1-s^2)$ normally present in the integration measure.

  While we  follow the JV work in spirit, we do not follow it in all technical details. The reason is we 
  intent to widen the scope of the study, to 3,4,5-quark sectors, and we need to keep
  uniformity of approach for all of them. 
  
  The main deviation is that we will ignore details of the transverse motion, and simply add the mean $ \langle H_\perp \rangle$ to hadronic masses, where appropriate. The reason for it is clear: one needs to keep the size of Hilbert space
  manageable. 
  
 Another modification is technical. The functional basis selected by JV was the eigenfunctions of the two first terms in the Hamiltonian.
  We do not find it either important or especially beneficial, since  full Hamiltonian is non-diagonal and is numerically diagonalized anyway.  
  
  Another provision, limited to two-quark systems, is parity of the wave function,
  which is good quantum number. So we select (in this section only!) the set of even 
  harmonics (polynomials of $z^2$).

 % from 2bodies.nb
 
 The set of functions is normalized, so that 
\be  \int ds J(s) \psi_{n_1}(s) \psi_{n_2}(s)=\delta_{n_1n_2}  \ee
and then calculate $< n_1 | H | n_2 >$  matrix for all three terms of the Hamiltonian.  in general, if quarks have different masses (as in $K,K^*$ mesons). 
  The confining term we borrowed from  \cite{Jia:2018ary} and use it in the form 
  \be H_{conf}=-{\kappa^4 \over M_q^2} {\partial \over \partial s} J(s)  {\partial \over \partial s}  \ee
 where $J(s)=(1-s^2)/4$ is the integration measure. Two derivatives over momenta are quantum-mechanical version
 of the (longitudinal) coordinate sqiared: so it can be called ``a longitudinal version of the harmonic confinement". The measure appearing 
 between the derivatives is there to ensure that this term is a Hermitian operator. 
 
 Quartic interaction we use, as explained already, is not some version of NJL operator, but rather its significant part,
 namely the topology-induced 't Hooft Lagrangian. As explained above, its chiral structure is such that
 it does not contribute to vector mesons (chiral structure $LL+RR$). It means that diagonalizing the first two terms of
 the Hamiltonian we already get predictions for their wave function. 
 
The value of the parameters we use, $M_q = .337 \, GeV , 
\kappa = .227 \, GeV$, are just borrowed from JV paper \cite{Jia:2018ary}  one gets the ground state mass $m_\rho \approx 700\, MeV$. It is 10\% lower than
the experimental and Jia-Vary value, as we neglected the confinement in transverse direction.
 
 For the ground $\rho$ state we have then the WF of the approximate polynomial form 
 given in Appendix.
% \be \psi_{\rho}(s) \approx -1.51306  + 10.6471 s^2 - \ee
%$$ 27.3622 s^4  + 30.0252 s^6  - $$
% $$11.8485 s^8 + ...$$
% %plotted in Fig.\ref{rho_wf}.
  Note that it has rather narrow peak near the symmetry point $z=0$ or $x_1=x_2$, and is very small near
 the kinematical boundaries.

% 
% \begin{figure}[htbp]
%\begin{center}
%\includegraphics[width=6cm]{rho_wf}
%\caption{default}
%\label{default}
%\end{center}
%\end{figure}

\begin{figure}[htbp]
\begin{center}
\includegraphics[width=4cm]{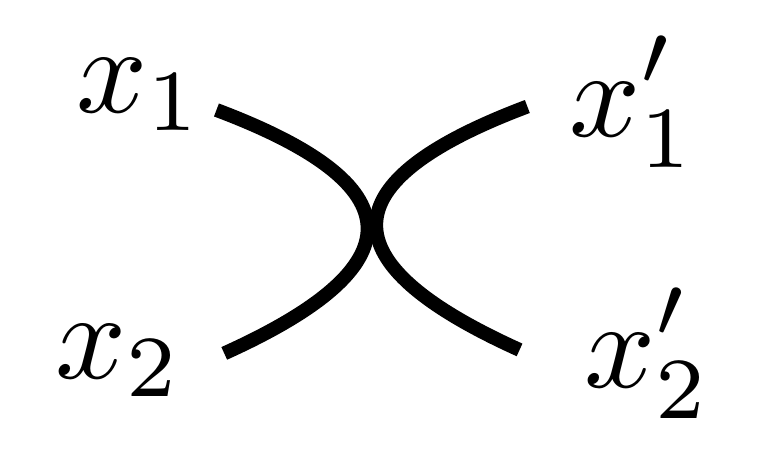}
\caption{default}
\label{fig_NJL_for_2}
\end{center}
\end{figure}

  Now we add the 4-fermion interaction, shown in Fig.\ref{fig_NJL_for_2}. Note that the left-hand-side and the right-hand-side include independent integrations, over $(x_1,x_2)$ in one state and over $(x_1',x_2')$ in the other. But since this gives total probability of $any$ values, the corresponding matrix is proportional to this curious matrix $ M_{all\,ones}$ which has $all$ matrix element being just 1:
\be H_{NJL} ^{n_1,n_2} = \pm G^2 M_{all\,ones} \ee
 $$ M_{all\,ones}^{n_1,n_2} \equiv 1 $$
The sign minus correspond to 4-fermion attraction in the pions, the sign plus to repulsion in the $\eta'$ channel. Let us explain why it is so, using the simplest 2-flavor case, in which 
$$|\pi^0 \rangle \sim {1 \over \sqrt{2}} (\bar u  u +\bar d d) $$  
and the isospin zero configuration orthogonal to it $\eta$ (becoming $\eta'$ in three-flavor theory) is 
$$|\eta \rangle \sim {1 \over \sqrt{2}} (\bar u  u -\bar d d) $$
In matrix elements of the Hamiltonian, $\langle \pi^0 | H |  \pi^0 \rangle$ and 
$\langle \eta | H |  \eta \rangle$, the nondiagonal operator $(\bar u u)(\bar d d)$ 
contribute with the opposite sign, making the pion light and $\eta$ heavy. The opposite sign
here is well known manifestation of explicit $U_a(1)$ breaking.

Adding this last term of the Hamiltonian to the first two, and diagonalizing, we obtain masses of these states.
The coupling we select from the requirement that pion be massless, which leads to mesonic
effective coupling $ G\approx 0.65\, GeV^2 $.
 
 The probability to find a quark at momentum fraction $x$, $P=x(1-x)\psi^2(x)$, should correspond to valence structure functions. (Of course, we still are in 2-quark sector and have no sea. There is no pQCD evolution and gluons. )   
 
\begin{figure}[h!]
\begin{center}
\includegraphics[width=6cm]{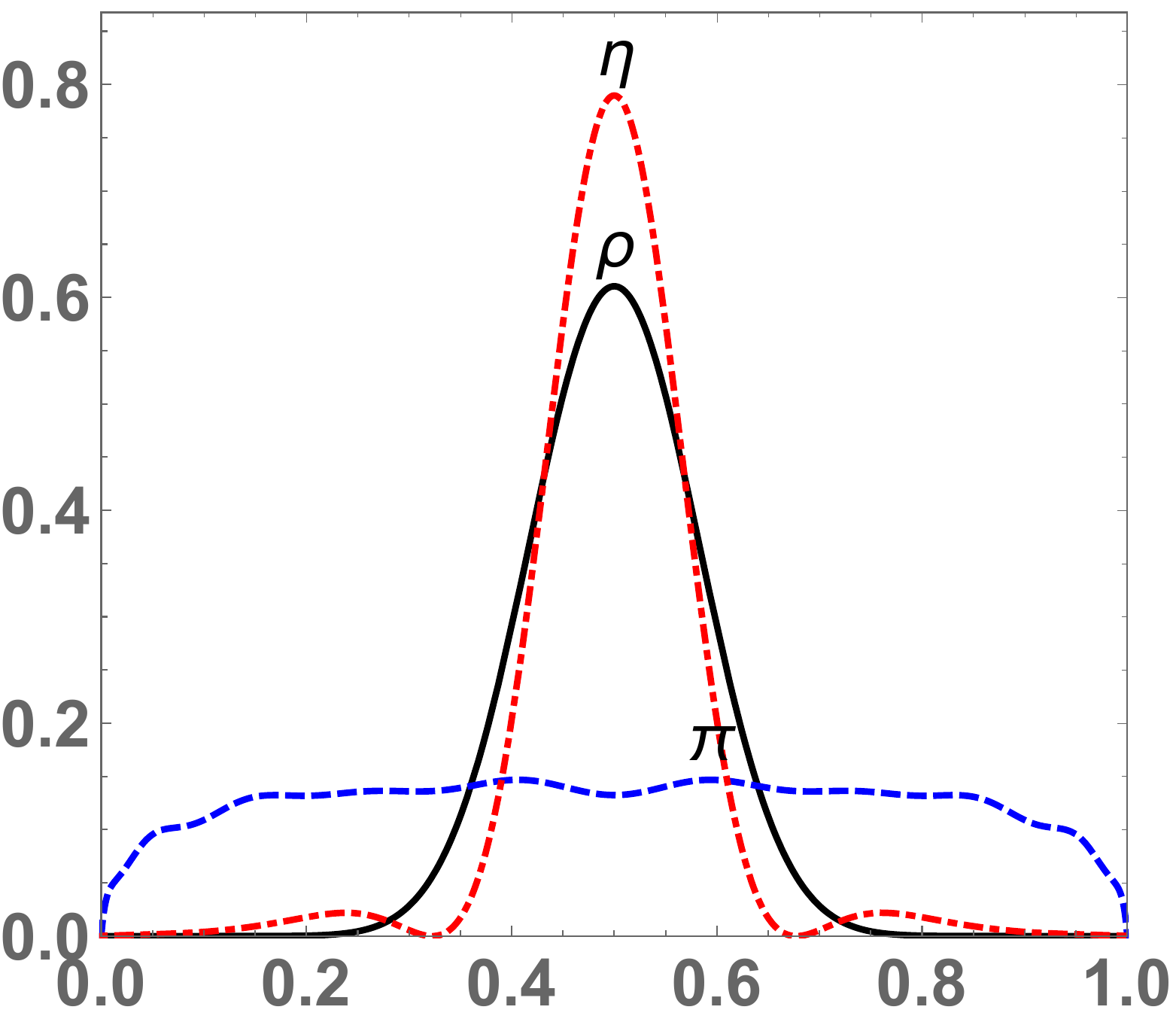}
\includegraphics[width=7.1cm]{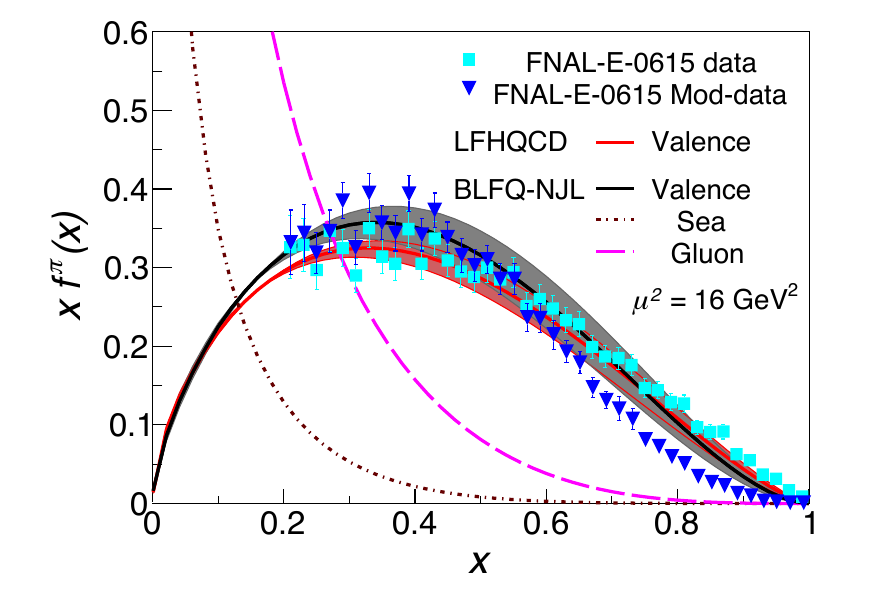}
\caption{Upper: momentum distribution for pion, rho and eta-prime mesons, calculated in the model. 
Lower (from \cite{Geesaman:2018ixo}) comparison between the measured pion PDF (points)  and the JV model (lines).}
\label{fig_pi_rho_eta}
\end{center}
\end{figure}

 The results from  the diagonalization of $H$, for the lowest states in the rho, pion and eta-prime channels are plotted in Fig.\ref{fig_pi_rho_eta}. Note that the predicted PDF for $\rho$ meson (in which the 4-fermion interaction is presumed to be absent) is peaked near the symmetric point $x=1/2,s=0$. The pion one, in contrast, has a completely different flat shape, . As one compares the lower plot to our result, one should keep in mind the fact that
 the PDF include also contribution from sectors with the quark number larger than 2,
 while ours (so far) do not. 
 
 Two historic remarks: (i)  one needs to mention an early model of the so called ``double-hump" pion wave function suggested in 1980's by 
 Chernyak and Zhitnitsky, in our notations corresponding to 
 the first harmonics $\psi_\pi(s)\sim s$. Our result 
 is not of course in agreement with it, but the 4-quark interaction does make the distribution
 much flatter, as compared to ``asymptotic" distribution from the measure $\sim x(1-x)=(1-s^2)/4$, corresponding to the zeroth harmonics $\psi_\pi(s)=const(s)$. \\(ii) Pion wave function was calculated within the instanton liquid model in \cite{Petrov:1998kg}: to the extent we can compare the results are similar.
 
 Note surprisingly, the PDF for ``repulsive" $\eta'$ channel moves in the opposite direction. In fact
 $two$ maxima are predicted, the larger one above $x=1/2$ and a small one near $x=1/4$. While these
 predictions is not possible to verify experimentally, one in principle should be able to compare it with
 lattice data and other models.

\section{Baryons as the three interacting quarks} \label{sec_3q}
In this pilot study we will pay little attention to quark/antiquark spin states.
Like transverse motion, it will be delegated to later studies with larger functional basis.
Let us just mention that 3-quark wave functions we will be discussing are, for Delta 
baryon 
\be | \Delta^{++} \rangle \sim \psi_\Delta(x_i)  |u^\uparrow(x_1) u^\uparrow(x_2) u^\uparrow(x_3)  \rangle \ee
and for the proton
%coefficient of the following quark states in the nucleon
\be | p \uparrow \rangle \sim \psi_p(x_i) \big( |u^\uparrow(x_1) u^\downarrow(x_2) d^\uparrow(x_3)  \rangle \label{def_p_wf} \ee
$$ - | u^\uparrow(x_1) d^\downarrow(x_2) u^\uparrow(x_3)  \rangle
\big) $$
and in what follows we will focus on the former component, in which the $d$ quark has the last momentum fraction $x_3$. For a review see \cite{Braun:2006hn}.

Let us just mention that some empirical information about
 3-quark LFWF of the nucleons come from exclusive processes. 
 Theory of these functions is, to our knowledge,  reduced to
 evaluation of certain moments, originally via QCD sum  rules and
 now from lattice studies. The 
calculation of the pQCD evolution as a function of $Q^2$, using the matrix of anomalous dimensions
for  leading twist operators was done in \cite{Braun:1999te}.
For doing so, it was important to use
%We will use
the so called conformal basis 
\be \Psi_{N,n}(x_i)\sim ({1+t })^n P_{N-n}^{(2n+3,1)}(-t) C^{3/2}_n(s) \label{eqn_conf_basis} 
\ee
 where $P_l^{(a,b)}(z)$ and $C^m_n(z)$ are Jacobi and Gegenbauer polynomials.the readers (Both are defined on $z\in [-1,1]$ and form the orthonormal basis, with
 the weight functions $(1-z)^a (1+z)^b$ and $(1-z^2)^{m-1/2}$
 , respectively.) 
 For consistency, will be using slightly different set of functions defined (\ref{eqn_conf_basis}) as orthogonal set. (We omit the discussion of normalization coefficients
of these functions, which are known but not particularly instructive.) 

The first physical effect we would like to incorporate is the part of the effective Hamiltonian coming from chiral symmetry breaking, the  constituent quark masses term
\be H_{mass}=M_q^2 \big( {1 \over x_1}+  {1 \over x_2}+{1 \over x_3}\big) \ee
%(if one of the quarks is strange, its mass value should be modified accordingly). 
The masses we use are the same as used in the fit for mesons in \cite{Jia:2018ary}.

The second included effect is confinement. 
We remind that since $x_i$ are momenta, the longitudinal coordinate are 
quantum conjugate to them, or $i\partial/\partial x_i$.
Making it as simple as possible, we follow 
what is done for mesons in \cite{Jia:2018ary}, we define this part of the Hamiltonian in the following
form, in $s,t$ coordinates
\be H_{conf}=-{\kappa^4\over J(s,t) M_q^2} \big[ {\partial \over  \partial s} J(s,t) {\partial \over  \partial s}+{\partial \over  \partial t} J(s,t) {\partial \over  \partial t}\big]
\ee
with the measure function $J(s,t)$  appearing in the $s,t,$ integration. Note that coefficient 4 in denominator is missing: this is cancelled by factor 4
coming from a difference between derivatives in $x$ and $s,t$ variables. 

The third (and the last) effect we incorporate in this work
is the topology-induced 4-quark interactions.  Note that topological 't Hooft Lagrangian
 is flavor antisymmetric. This means that it does not operate e.g. in 
 baryons made of the same flavor quarks, like the $\Delta^{++} =uuu$.
 Another reason why the 't Hooft vertex should be absent is in any states in which all chiralities of quarks are the same, like $LLL $. For both these reasons, $\Delta^{++}$ is not affected by topology effects,
  therefore serving as a  benchmark (like the $\rho$ meson did in the
 previous sections). 

Thus, we have a prediction for the masses of  $\Delta$ baryons and their excited states,
as well as the corresponding LFWFs.  Using the quark mass and confining parameter same as mesons,
we get the mass squared of the lowest 3-quark state to be $1.28\, GeV^2$. Following what
we learned from mesonic case about transverse energy correction, we add it twice and get
\be M_\Delta =\sqrt{1.28+2\times .12}= 1.23 \, GeV \ee
in perfect agreement with the experimental Bright-Wigner mass $M_\Delta =1.230-1.234\, GeV$

Couple of other comments at this point: \\(i)
Without confining term, the  mass squared of the lowest 3-quark state is only slightly smaller,
$1.11\, GeV^2$. This is because this state is dominated by the lowest harmonics,
as we will show below, and the Laplacian in it is zero. In other words, the predicted LFWS
of Delta is rather flat.\\
(ii) The  mass squared of the next state is $1.679\, GeV^2$, and with a transverse correction it leads to
\be M_\Delta =\sqrt{1.679+2\times .12}= 1.39 \, GeV \ee
Experimentally, the second $\Delta'$ state has Bright-Wigner mass higher, at $1.5-1.7\, GeV$.
We generally  do not expect the model to adequately describe the excited states,
since transverse degrees of freedom are not really represented in the WF.

\begin{figure}[htbp]
\begin{center}
\includegraphics[width=5cm]{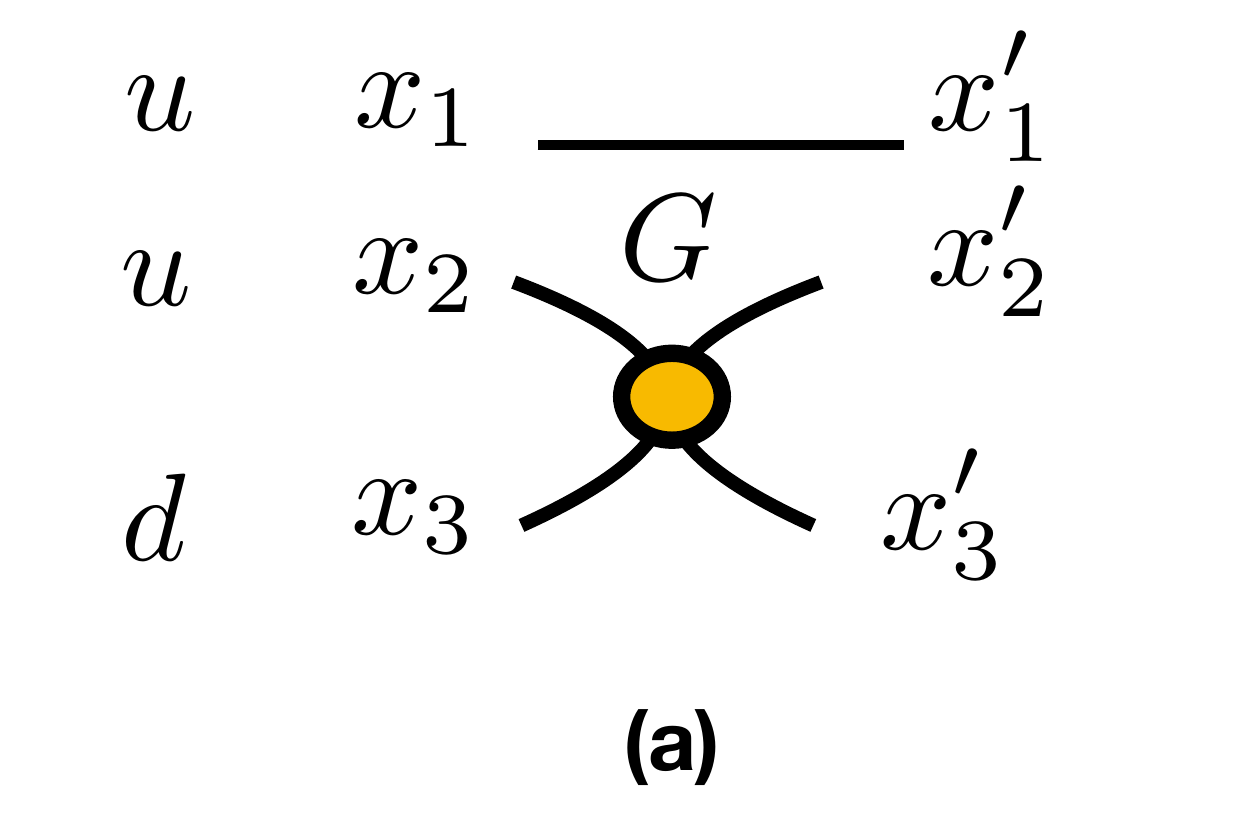}
\includegraphics[width=4cm]{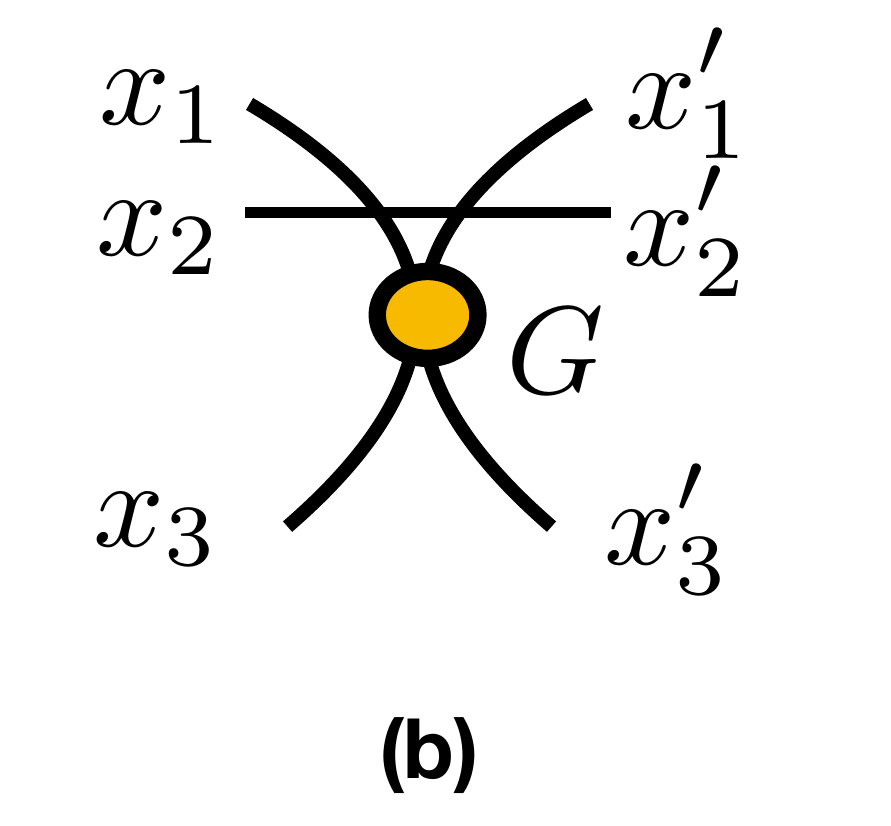}
\caption{Two terms in the Hamiltonian, describing rescattering in $ud$ channel due to residual local 4-quark interaction}
\label{fig_G_in_N}
\end{center}
\end{figure}

 The next step is implementation of 4-quark residual interaction. 
In the notations used (\ref{def_p_wf}),  we focus on the former proton state, in which $d$ quark has momentum fraction $x_3$. Unlike general NJL operators, the topology-induced 
one is flavor antisymmetric, acting in $(ud)$ channel only. 
   So, in principle there can be
 diagrams (a) and (b), shown in Fig.\ref{fig_G_in_N}, with interaction
  in (2,3) and (1-3) pairs, but none in $uu$
pair due to Pauli principle for the quark zero modes. 
 Furthermore, the topology-induced NJL-like local attractive term only appear in the  
 spin-zero diquark channel, and therefore spin directions of $u$ and $d$ quarks must be opposite. This eliminates the diagram (b).  Note also, that two $u$ quarks are distinguishable as they have opposite spins: thus there is no exchange diagrams.

The kinematics of the diagram is  as follows.
In general, a  transition amplitude   from initial set $x_1,x_2,x_3$ of kinematical variables to the final ones
 $x_1',x_2',x_3'$ include 4 integrations. However, since one of the quarks
 does not participate, there is always one delta function for non-interacting quark (e.g. for (a) interacting, $x_1=x_1'$, and the matrix elements of the (a,b) diagrams include effectively 3D integrals. Yet since all the functions and kinematical boundaries are already defined in terms of $s,t,s',t'$, we preferred to keep full 4-d integrations, implementing delta function approximately, by narrow  Gaussian.

The Hamiltonian matrix for residual interaction we write in the form, for diagram (a)
\be \langle i_1 | H_{a} | i_2 \rangle = -G\int ds dt J(s,t) \psi_{i_1}(s,t)^2 \ee
$$\times \int ds' dt' J(s',t') \psi_{i_2}(s',t')^2 \delta\big(x_1(s,t)-x_1'(s',t')\big)
$$
 The sign minus corresponds to attraction in
 the $ud$ channel. As $G$ increases, the mass of the lowest state decreases: 
we stop when it becomes equal to the nucleon mass. This allows us to fix our first
model parameter \be G\approx 17 \,\,\, GeV^2 .\ee 

(We already noticed in the pion case, that an attractive 4-fermion interaction in
principle makes the vacuum unstable and create the massless pion. At chosen $G$ there are signs of this phenomenon, as
 imaginary parts of squared masses of higher nucleon states. The nucleon's mass 
remains real and far from zero.)

Now we are ready to explore properties of the obtained  nucleon wave function,
which one can read off from eigenvector of the Hamiltonian.

\begin{figure}[h!]
\begin{center}
\includegraphics[width=6cm]{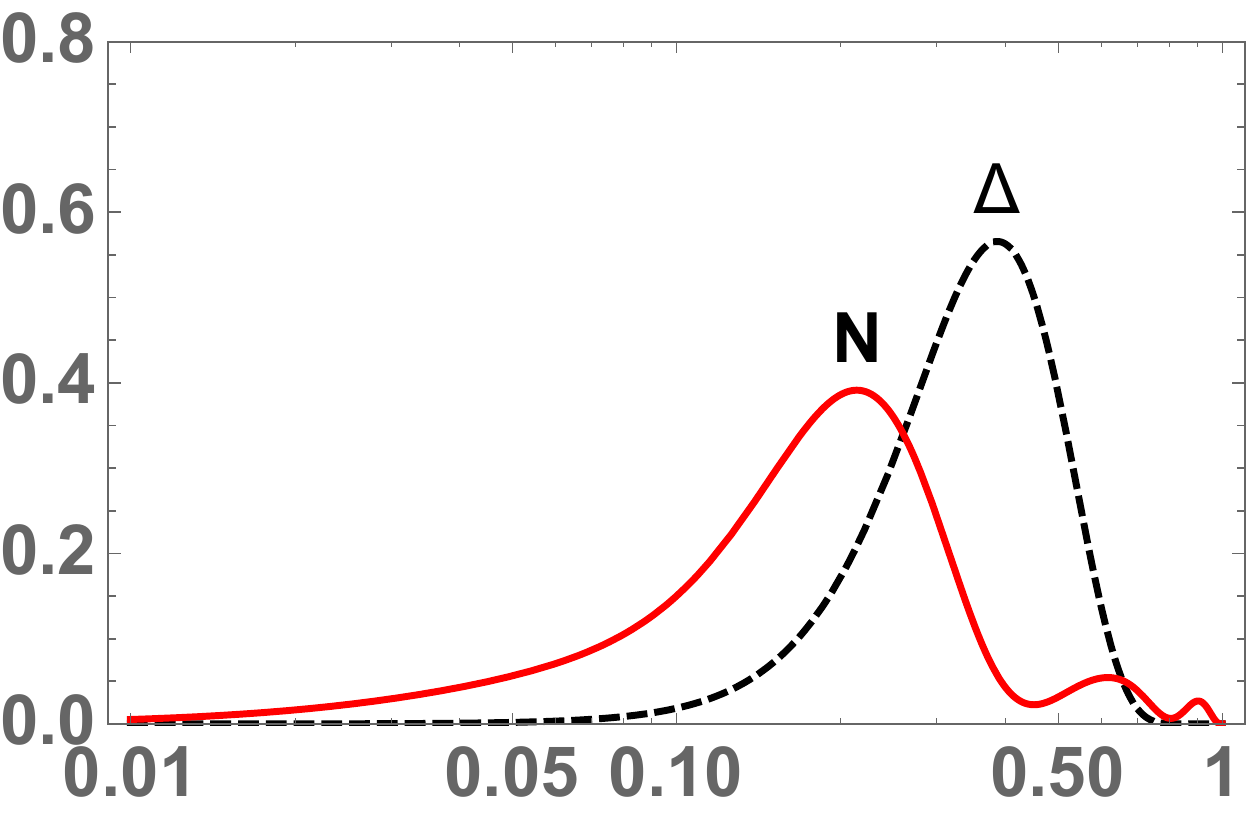}
\includegraphics[width=6.3cm]{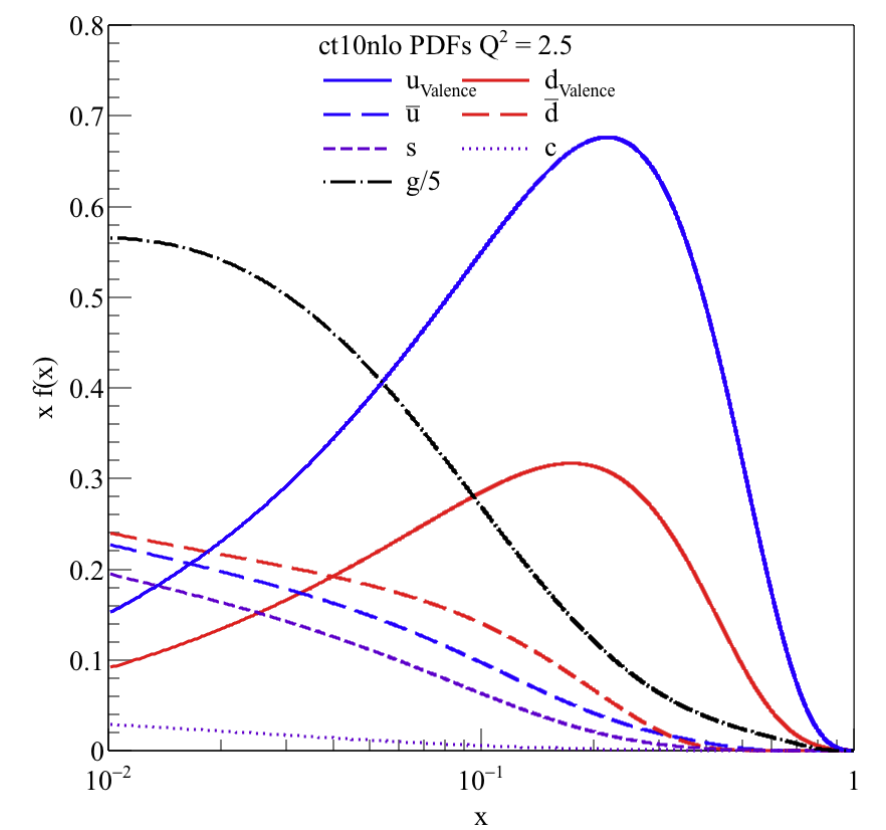}
\caption{
Upper: our calculation of the $d$ quark distribution in the Nucleon times $x$, $x d(x)$ (red, solid)
and Delta (black, dashed) states. For comparison, the lower plot shows empirical structure functions
(copied from \cite{Geesaman:2018ixo} ), where the valence $xd(x)$ is also shown in red. }
\label{fig_d_structure}
\end{center}
\end{figure}

The simplest quantity to display are single-quark densities, corresponding to nucleon PDFs.
In particular, the distribution over $x_3$, or the $d$ quark, depends in our notations
only on variable $t$. So, 
 performing integration over variable $s$ keeping $t$ arbitrary yields the d-quark distribution
\be d(x_3=1-2t)=\int_{-1}^1 ds J(s,t) \Psi_N^2(s,t)
\ee 
This   distribution, multiplied by $x$ is shown in Fig.\ref{fig_d_structure}(upper), for Delta and Nucleon
wave functions we obtained. Two comments: (i) the peak in the Nucleon distribution moves to lower $x_d$, as compared to $x=1/3$ expected in $\Delta$ and
non-interacting three quarks; (ii)  there appears larger tail toward small $x_d$ in the nucleon,
but also some peaks at large $x_d$
Both are unmistakably the result of local $ud$ pairing (strong rescattering) in a diquark cluster.

The Fig.\ref{fig_d_structure} (lower), shown for comparison, includes the empirical valence $xd(x)$  distribution, also shown by red solid line. The location of the peak (i) roughly corresponds to data, and (ii)
the presence of small-$x_d$ tail is also well seen (recall that what is plotted is
distribution times $x$). The experimental distribution is of course
much smoother than ours.
 It is expected feature: our wave function 
 is expected to be ``below the pQCD effects", at resolution say $Q^2 \sim 1\, GeV^2$, while
 the lower plot is at $Q^2=2.5\, GeV^2$, ant it includes certain amount of pQCD gluon radiation.
 contains higher order correlations between quarks. 
 
 Our distribution displays certain structure, related to quark-diquark component
 of the wave function, which  may or may not be true. It
 is definitely present in the model used. As for the data, it is not so obvious: collaborations 
 fitting the data imply
 simple functional forms, so their smoothness of the empirical PDFs 
also may be questioned.

In order to reveal the structure, 
one can of course compare the wave functions without any integrations, as they depend
on  two variables $s,t$ only.  Such plots, of $J(s,t) \Psi^2(s,t) $, we show  In Fig.\ref{fig_Delta_N_CZ}  for 
the Delta, the nucleon, and some model discussed  in \cite{Chernyak:1987nv} (see Appendix). While the Delta
shows a peak near the symmetry point $x_1=x_2=x_3$ as expected, without any other structures, 
our Nucleon WF indicate more complicated dynamics. Indeed, there appear several bumps, most prominent near $s\approx 1,t\approx 1$ which is $x_1\approx 1$. Apparantly, it is about the same peak location which CZ
wave functions wanted to emulate. Such strong peaking corresponds to large momentum transfer inside the $ud$ diquark clusters. Yet there is also the peak in the middle, roughly
corresponding to that in Delta. So, the nucleon wave function is a certain coherent mixture of a three-quark 
and quark-diquark components.

\begin{figure}[h!]
\begin{center}
\includegraphics[width=5cm]{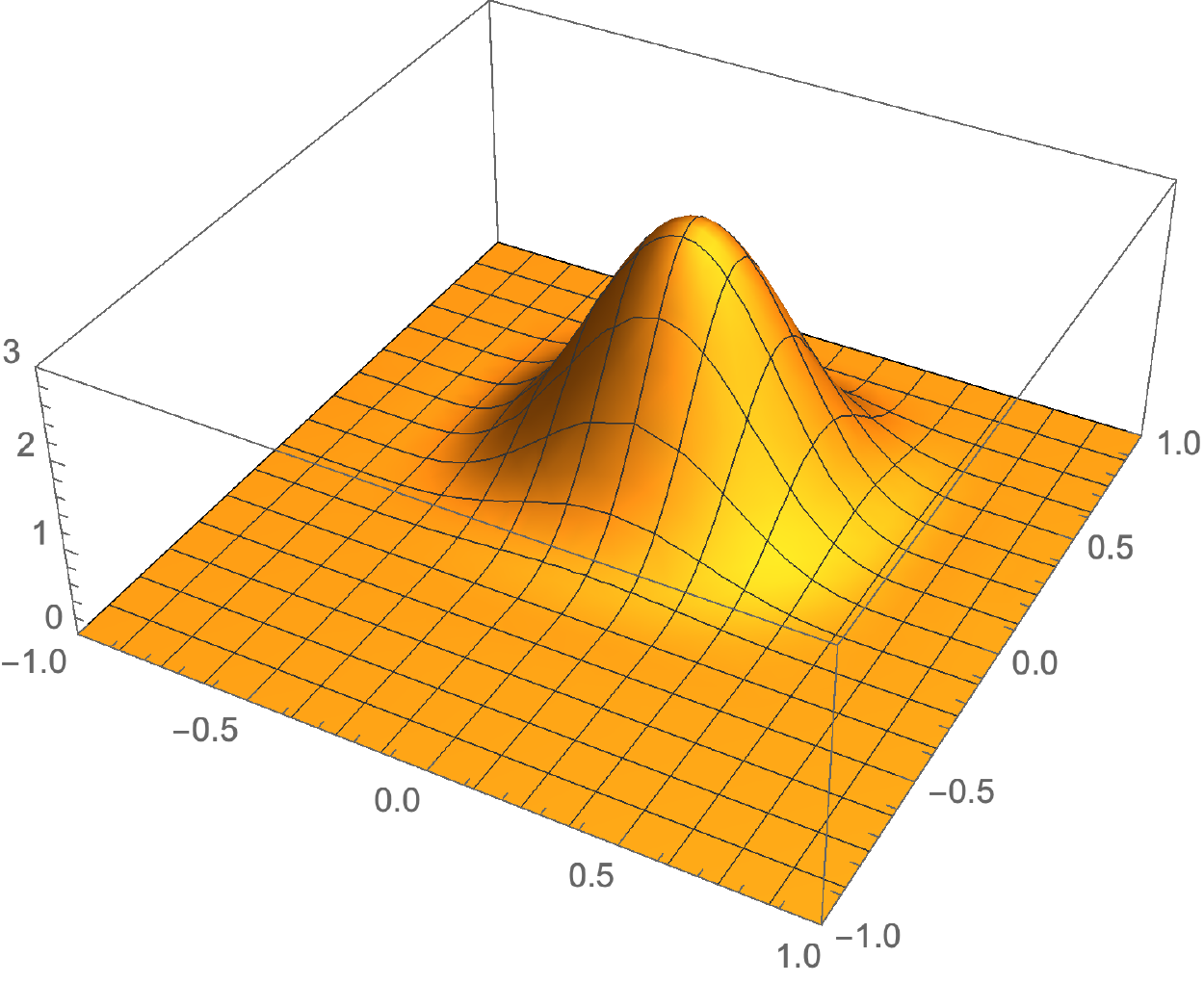}\\
\includegraphics[width=5cm]{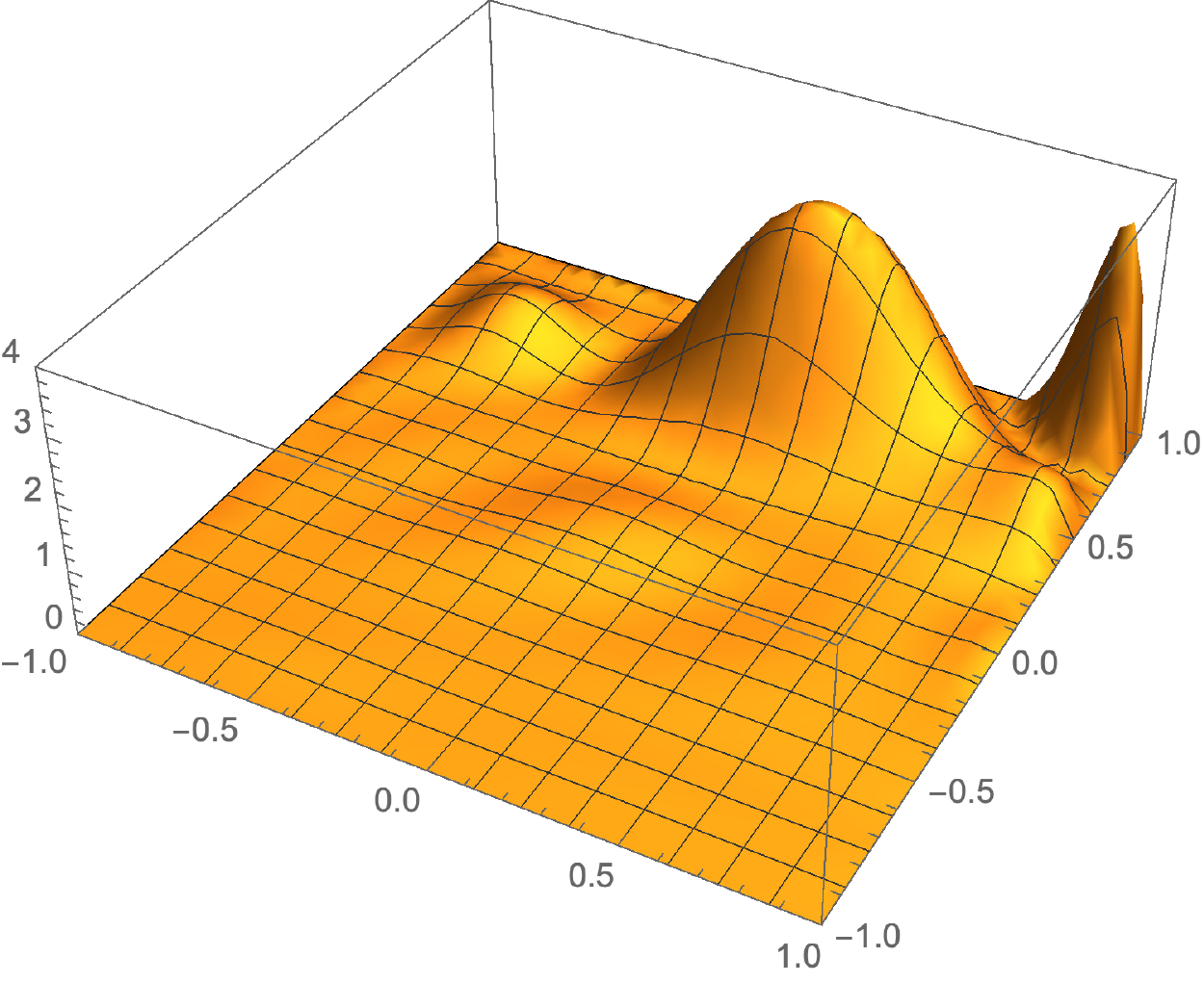}\\
\includegraphics[width=5cm]{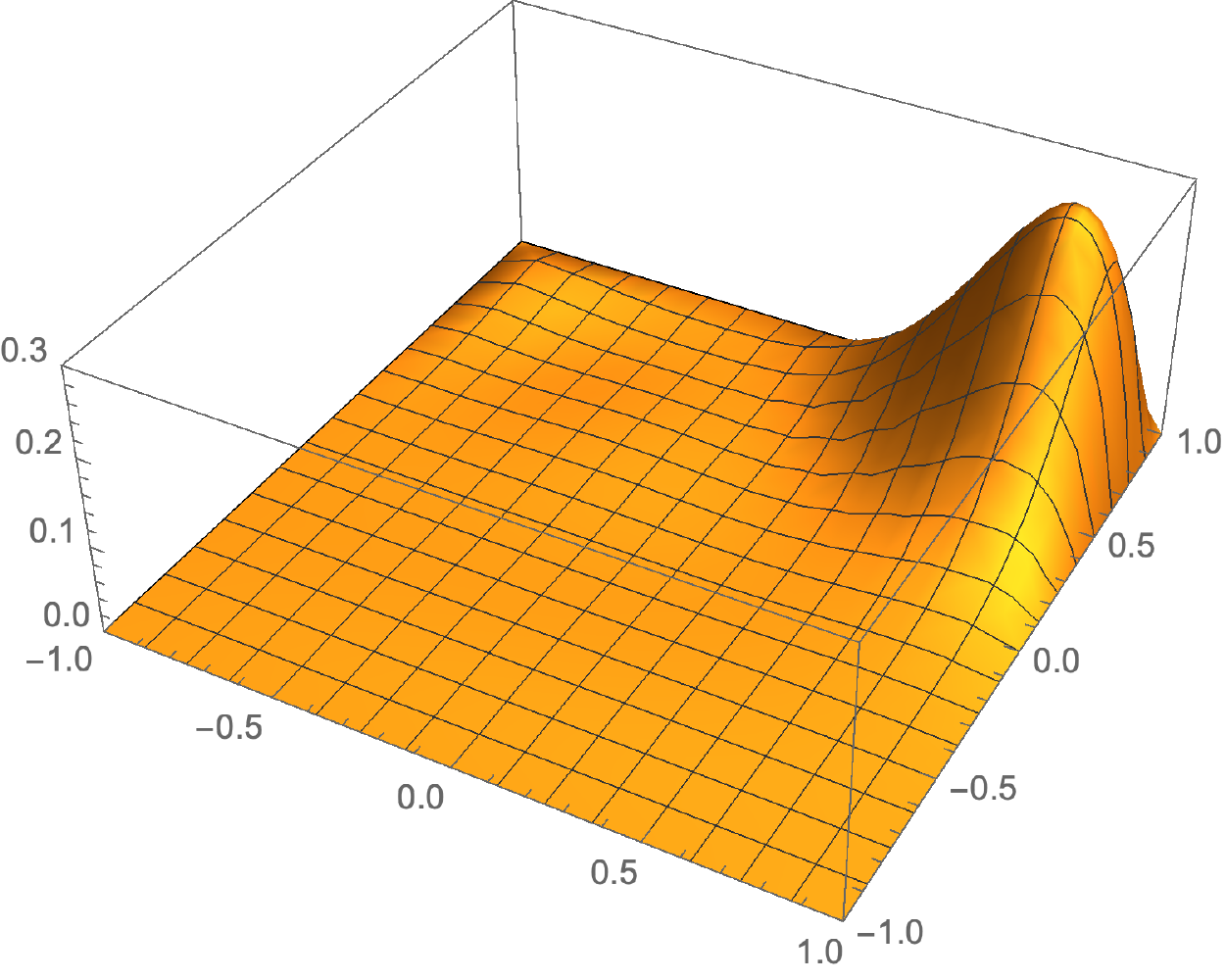}
\caption{Two upper plots show the probability distribution $J(s,t)\Psi^2(s,t)$ in $s,t$ variables, for the Delta
and Nucleon lowest states, as calculated from the model. The lower plot corresponds to Chernyak-Zhitnitsky model wave function $N_{IJ}$ (\ref{CS_wfs}) , shown for comparison.
 }
\label{fig_Delta_N_CZ}
\end{center}
\end{figure}

Additional information about the nucleon wave function may be obtained from
the amplitudes corresponding to particular basis states.  We remind that its definition is
in (\ref{3part_basis} ) (up to normalization, which we fixed from $1=\int ds dt J(s,t) \psi_{n,l}(s,t)^2$ ).

\begin{figure}[h!]
\begin{center}
\includegraphics[width=5cm]{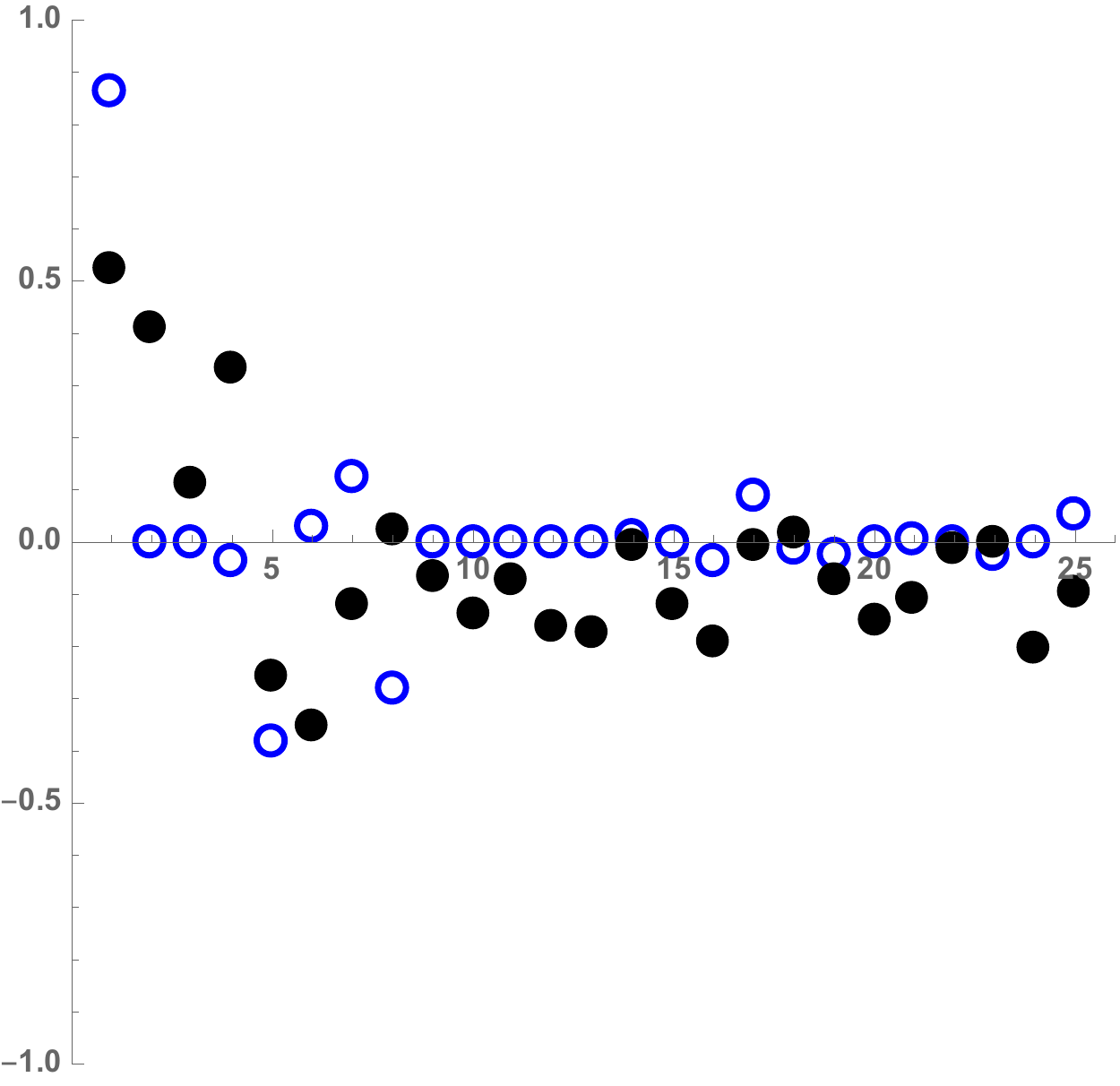}
\includegraphics[width=5cm]{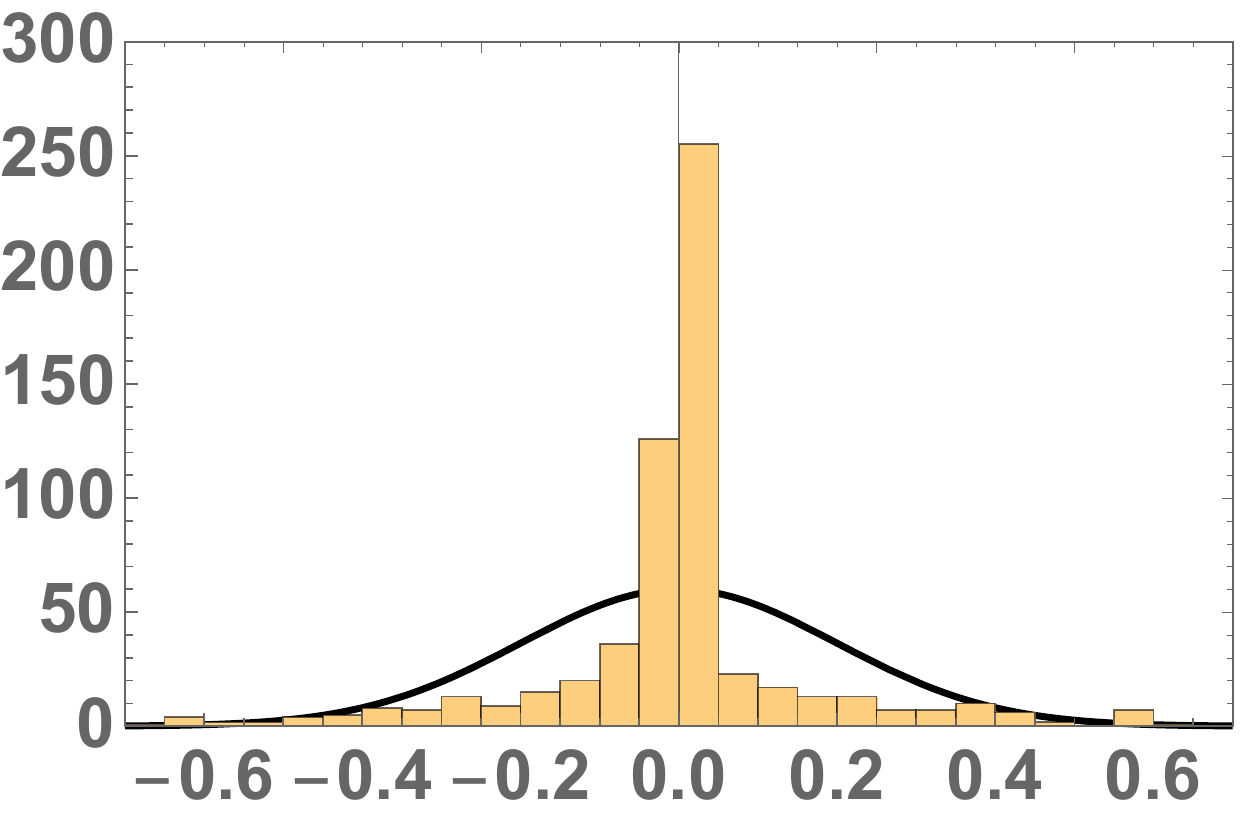}
\includegraphics[width=5cm]{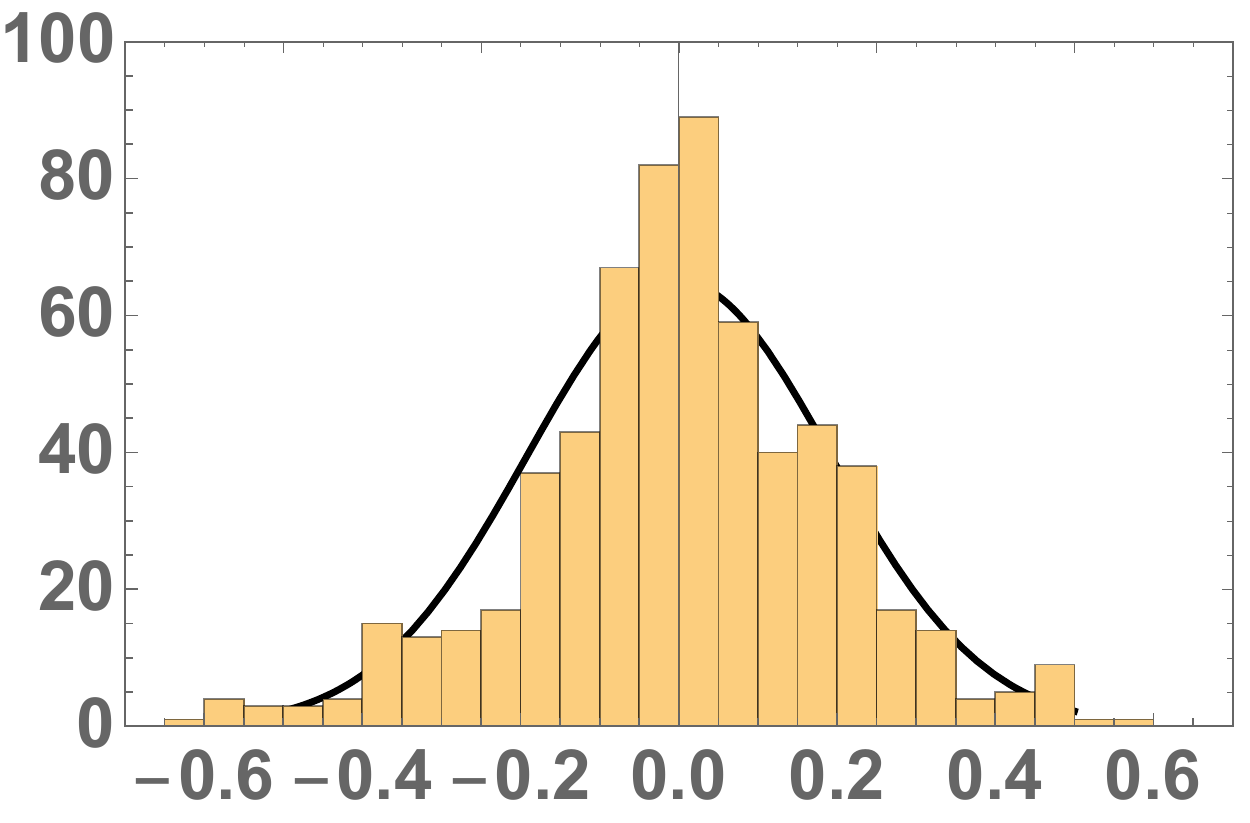}
\caption{Upper: The coefficients of the wave function $C^\alpha_n $ (the order of basis functions is specified in Appendix), for the ground state Delta (open points) and the Nucleon (closed points) states. Two lower plots contain histograms of these coefficients for 25 lowest
 Delta and Nucleon states, respectively. The black lines in the background are the Gaussians with
 the appropriate width. }
\label{fig_harmonics}
\end{center}
\end{figure}

The wave function coefficients $C^\alpha_n $, 
defining expansion in the basis functions
$$ | \alpha >=\sum_n C^\alpha_n | n >$$ 
  are shown in Fig. \ref{fig_harmonics}.
  The upper plot compares those for the ground state
    Delta and Nucleon channels.
Note first that the largest coefficients are the first (corresponding to trivial $\psi_{0,0}(s,t)\sim const(s,t)$). Furthermore, for the Delta it is close to one, while it is only $\sim 1/2$ for the nucleon. The fraction of ``significant coefficients" is much larger  for $N$.   The nucleon wave function has a nontrivial tail toward  
 higher $n,l$ harmonics
which does not show any  decreasing trend.
One may in fact  conclude that  convergence of the harmonic expansion is not there. This 
 can be traced to apparent peak  near $x=1$, perhaps the pointlike residual interaction leads
 to true singularity there. 

Two lower plots of  Fig. \ref{fig_harmonics} address the distribution of the wave function coefficients $C^\alpha_n $ in these two channels, without and with the residual 4-quark interaction. It includes not the ground state but the lowest 25 states in each channel.
As it is clear from these plot, in the former (Delta) case the distribution is 
very non-Gaussian, with majority of coefficients being small. The latter (Nucleon) 
case, on the other hand, is in agreement with Gaussian. In other quantum systems,
e.g. atoms and nuclei, Gaussian distribution of the wave function coefficients $C^\alpha_n $  is usually taken as 
a  manifestation of ``quantum chaos". In this language, we conclude that
our model calculation shows that  the residual 4-quark interaction
 leads to chaotic motion of quarks, at least
inside the  Nucleon resonances. (If this conclusion surprises the reader,  we remind that
the same interaction was shown to  produce chaotic quark condensate in vacuum. In particular,
numerical studies of Interacting Instanton Liquid in vacuum has lead to 
Chiral Random Matrix theory of the vacuum Dirac eigenstates near zero,
accurately confirmed by lattice studies.)  

\section{Pentaquarks} \label{sec_5q}
Jumping over the 4-quark LFWFs we proceed to the next sector, with $qqqq\bar q$ composition.  Although
the main intension is ``unquenching" our quark model, coupling sectors with different number of quarks,
together in common Hamiltonian eigenstate, we start in this section with a study of the pentaquark
$qqqq\bar q$ states as such. One can also view this section as calculation of the Hamiltonian matrix in the 5q sector, to be coupled with the 3q one in the next section. For simplicity, we 
assume the four quarks to be in 4 distinguishable spin-flavor states, namely $u^\uparrow u^\downarrow d^\uparrow d^\downarrow$, which  minimizes the Fermi repulsion effects by eliminating the exchange diagrams.

There is no change in the model or its parameters: all we do is writing the Hamiltonian as a matrix in a certain basis and diagonalizing it. Four kinematical variables $z_i=s,t,u,w$, the integration measure $J(s,t,u,w)$ and
the set of orthogonal functions, were already defined in section \ref{sec_kinematics}. Presence of 4 integers, numbering
the degree of the corresponding polynomials, naturally restricts our progress toward large values of each. Matrix elements of the Hamiltonian become in this sector 4-dimensional integrals. Since this is a pilot study, restricted to laptop-based calculations
inside Mathematica, we evaluate $\sim 10^3$ such matrix elements. The set of states $(l,m,n,k)$ used in this section contains the following 35 states  for which $l+m+n+k<4$ 
listed in Appendix
%$$
%(0, 0, 0, 0),(1, 0, 0, 0),(0, 1, 0, 0), (0, 0, 1, 0), (0, 0, 0, 1), $$
%$$  (1, 1, 0, 0), (1, 0, 1, 0), (1, 0, 0, 1), (0, 1, 1, 0), (0, 1, 0,    1), (0, 0, 1, 1), $$
%$$  (2, 0, 0, 0), (0, 2, 0, 0), (0, 0, 2, 0), (0, 0, 0, 2),
%  (0, 1, 1, 1), (1, 0, 1, 1), $$ $$  (1, 1, 0, 1), (1, 1, 1, 0), 
%  (2, 1, 0, 0), (2, 0, 1, 0), (2, 0, 0, 1), (1, 2, 0, 0), $$ $$ 
%  (0, 2, 1, 0), (0, 2, 0, 1),  (1, 0, 2, 0), (0, 1, 2, 0), (0, 0, 2, 1), 
% (1, 0, 0, 2), $$ $$ (0, 1, 0, 2), (0, 0, 1, 2),
%  (3, 0, 0, 0), (0, 3, 0, 0), (0, 0, 3, 0), (0, 0, 0, 3)
%$$

Matrix elements of the quark-mass term and the (longitudinal) confining terms, $H_{mass}=M_q^2\sum_{A=1,5} (1/x_A)$ and $H_{conf}=-(\kappa^4/J/M_q^2) (\sum_{i=1,4} \partial_{z_i} J \partial_{z_i})$ are calculated and 
diagonalized. Let us comment on some of the results.

The lowest eigenvalue was found to be $M^2_{min.penta}=4.04 \, GeV^2$. Following the procedure we 
adopted for mesons, we add transverse motion term (treated as constant) $4\times 0 .12 = 0.48 \, Gev^2$.
Taking the square root, the predicted lowest mass of the light-flavor pentaquark (due to chiral mass and confinement) is therefore 
\be  M_{min.penta}=2.13 \, GeV \ee
To get this number in perspective, let us briefly remind the history of the light pentaquark search.
In 2003 LEPS group reported pentaquark $\Theta^+=u^2d^2\bar s$  with surprisingly  light mass, of only $1.54 \,GeV$,
$0.6 \, GeV$ lower than our calculation (and many others) yield. 

Of course, so far the residual perturbative and NJL-type forces were not included. 
Quick estimates of the time (including mine \cite{Shuryak:2003zi})  suggested that 
since $ud$ diquark has binding energy of $\Delta M_{ud}\approx -0.3\, GeV$ and the pentaquark candidate has two of them, one gets to ``then observed" mass of  $1.54 \,GeV$. 

Several other experiments were also quick to report observation of this state, till
other experiments (with better detectors and much high statistics) show this pentaquark candidate does not really exist. 
 Similar sad experimental status persists for all  6-light-quark dibaryons
, including the flavor symmetric $u^2d^2s^2$ spin-0 state much discussed in some theory papers. 

A lesson for theorists is, as often, not to trust any simple estimates but proceed to some
more consistent calculation. In the framework of instanton liquid model student Pertot and myself studied diquark-diquark
effective forces, by correlating pentaquark operators in standard Euclidean setting. 
We found  \cite{Shuryak:2005pk}
 that, like for baryons,  consistent account for Fermi statistics generates strong repulsion between diquarks,
 basically cancelling the presumed attraction.  Same trend was observed in
 multiple lattice studies, e.g. by the Budapest group \cite{Csikor:2006jy}:
 no pentaquark states close to  $1.54 \,GeV$ in fact exist, while all physical states   are much heavier. 

Returning to our calculation, few more comments. First,  about the role of the confining
term of the Hamiltonian: without it, the lowest eigenvalue is at $3.82\, GeV^2$, and with it it is (as reproted above) $4.04 \, GeV^2$. The difference due to longitudinal confinement thus is $0.22\, GeV^2$. Twice this
value (because there are two transverse direction) gives $0.44\, GeV^2$, well consistent with 
 our empirical correction due to transverse motion of $4 \times 0.12=0.48 \, GeV^2$. 
 
 The next-to-lowest eigenstate of the Hamiltonian has the mass $4.87 \, GeV^2$, demonstrating rather large
 gap with the lowest state. We have so far made no investigation of why does it happen in this case, but not the others. 
 
 Even at the current level of approximation, without residual 4-fermion interactions, the  pentaquark wave function of the lightest eigenstate turned out to be rather nontrivial. Since it is a function of 4 variables, there is
 no simple way of plotting it. 
 Suppressing a bunch of small terms, 
 it can be written explicitly as polynomial, see Appendix.
 
\begin{figure}[h!]
\begin{center}
\includegraphics[width=5cm]{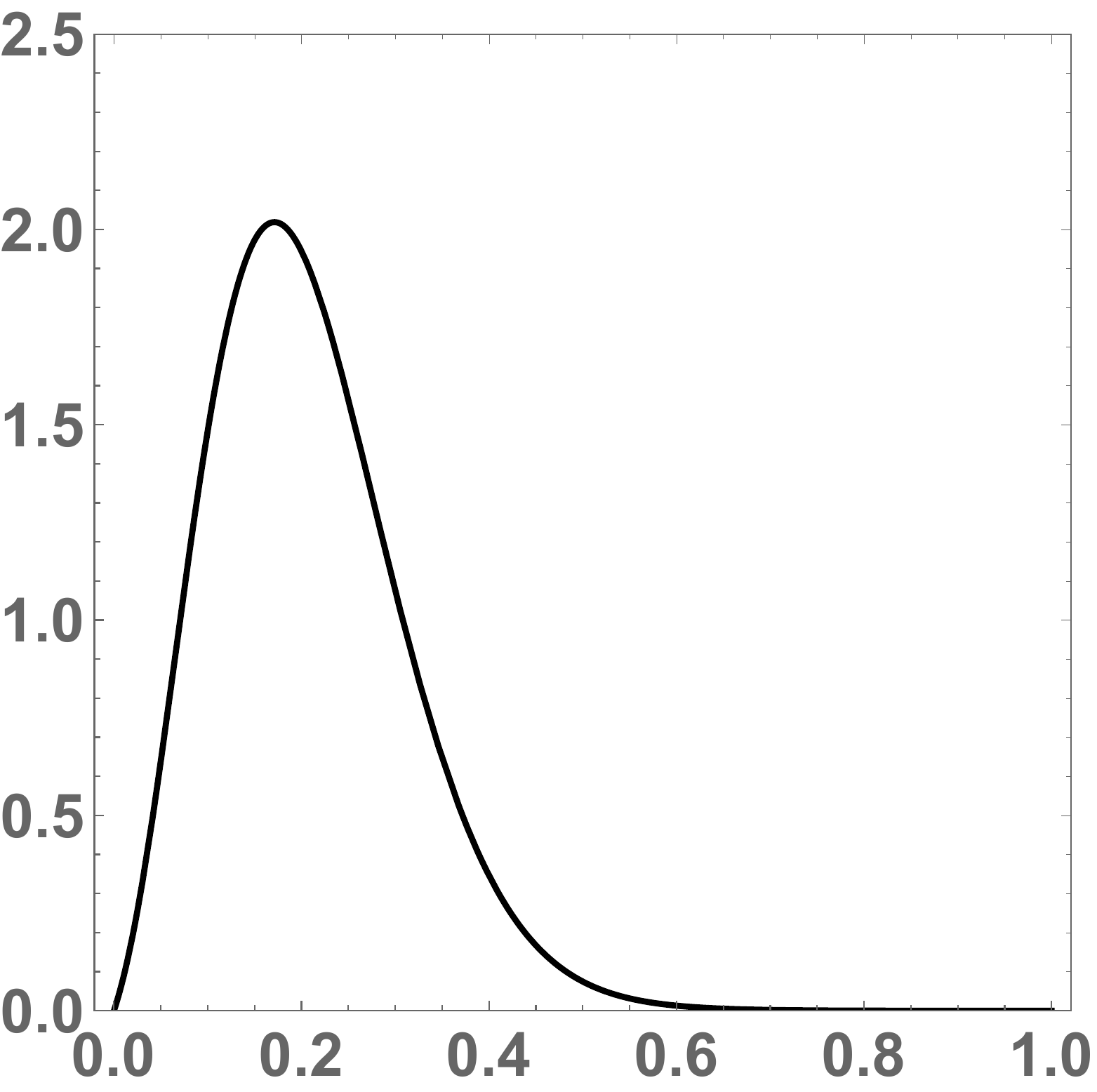}
\caption{Probability $P(X)$ to find a quark with momentum fraction $X$
in the lowest pentaquark state, calculated from the wave function given in the Appendix. Note that in this WF the
residual 4-quark interaction has not been yet implemented.}
\label{fig_penta_1body}
\end{center}
\end{figure}

Note that the coefficients of highest  order terms are not small: and in this sense 
the set of state was insufficient to show convergence.  While there are nontrivial correlations between variables,  a single-body distribution
calculated from it does not exhibit anything unexpected, see Fig.\ref{fig_penta_1body}.

\section{The 5-quark sector of the baryons } \label{sec_3q_5q}
%\subsection{Is the 5-quark sector small?}
As already mention in Introduction, the initial motivation for this work was
precisely a development of  framework in which one can consistently calculate
the wave functions in the lowest sector (three-quark for baryons), relating
it to the observed parameters of the {\em exclusive processes} (formfactors etc)  and the {\em valence quark PDFs},
and the proceed with calculation of the other sectors of the physical state.

\begin{figure}[htbp]
\begin{center}
\includegraphics[width=5cm]{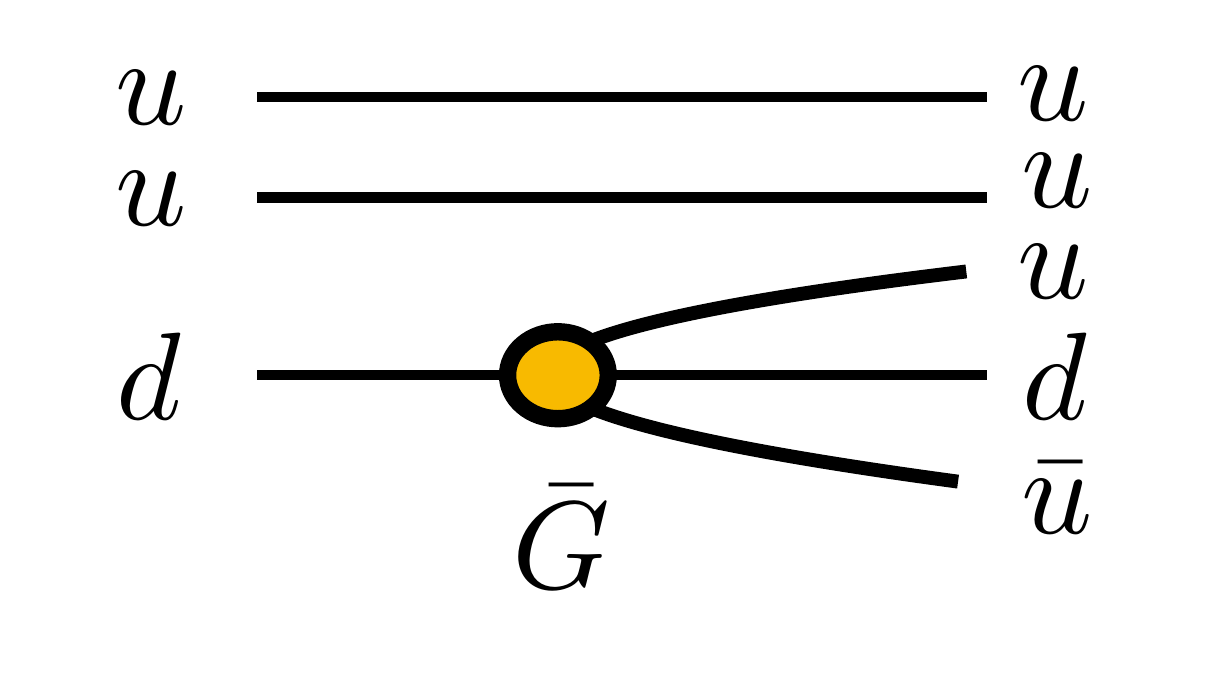}
\caption{The only diagram
in which 4-quark interaction connects the 3 and 5 quark sectors, generating
 the $\bar u$ sea.}
\label{fig_3_to_5}
\end{center}
\end{figure}

\begin{figure}[h!]
\begin{center}
\includegraphics[width=6cm]{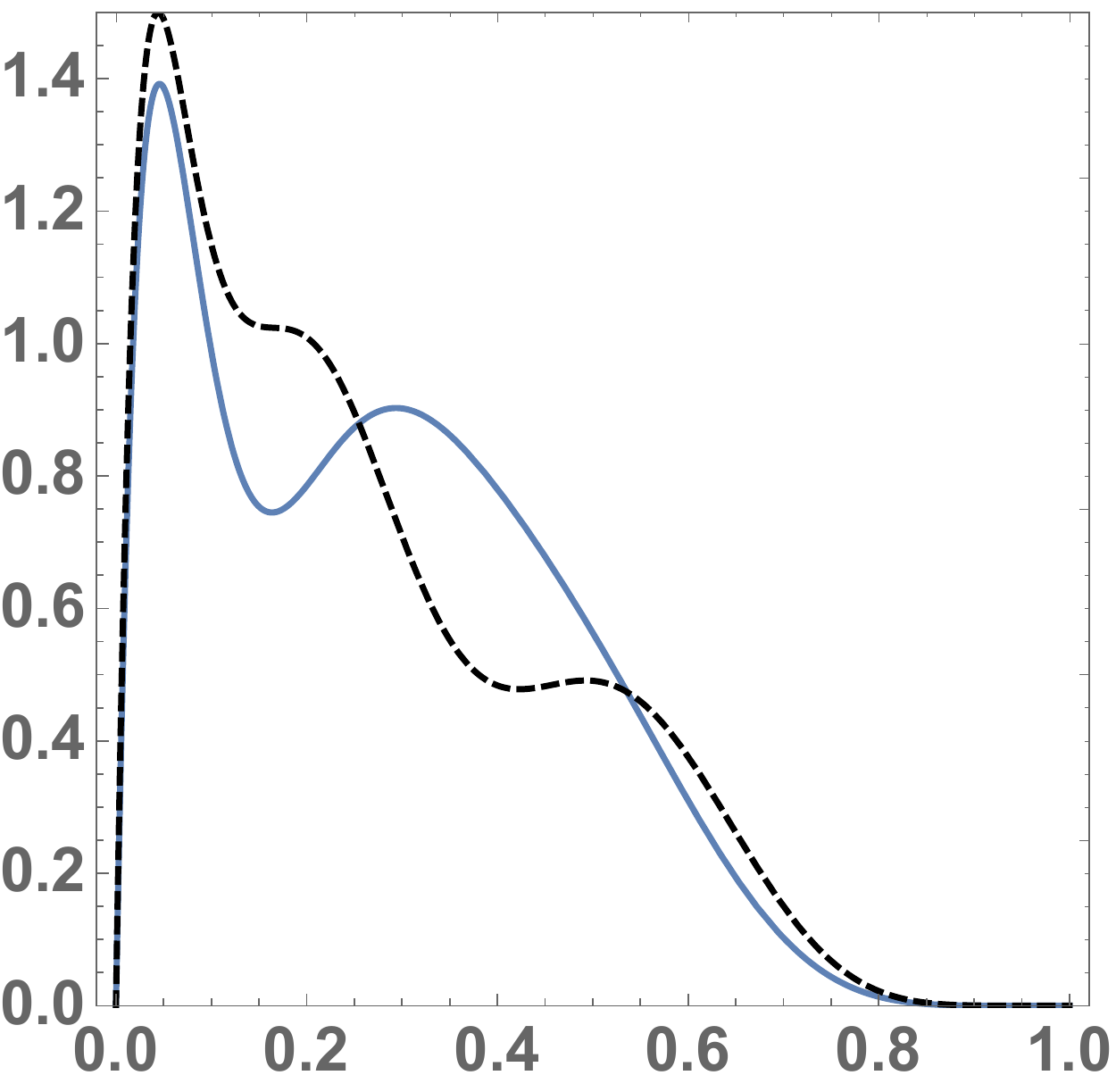}
\caption{The distribution over that $\bar u$ in its momentum fraction, for the
Nucleon and Delta 5-quark ``tails" (solid and dashed, respectively).}
\label{fig_antiu}
\end{center}
\end{figure}

The Hamitonian matrix element corresponding to the diagram shown in Fig\ref{fig_3_to_5}
we calculated between the nucleon and each of the pentaquark wave functions, defined above, 
by the following 2+4 dimensional integral
over variables in 3q and 5q sectors, related by certain delta functions
 \be \langle N | H | 5q,i  \rangle=\bar G \int ds dt J(s,t) ds' dt' du' dw' J(s',t',u',w') \ee
$$ \psi_N(s,t) \delta(x_1-x_1') \delta(x_2-x_2') 
 \psi_i (s',t',u',w') $$
The meaning of the delta functions is clear from the diagram, they are of course expressed via proper integration variables and  numerically
approximated by narrow Gaussians. After these matrix elements are calculated, the
5-quark ``tail" wave function is calculated via standard perturbation theory expression 
\be \psi_{tail}(s',t',u',w')= -\sum_i { \langle N | H | 5q,i \rangle \over M_i^2-M_N^2} \psi_i (s',t',u',w')  \label{eqn_tail} \ee
The typical value of the overlap integral itself for different pentaquark state is $\sim 10^{-3}$,
and using for effective coupling $\bar G$ the same value as we defined for $G$ 
from the nucleon, namely $\sim 17 GeV^2$, one finds that admixture of several pentaquarks
to the nucleon is at the level of a percent.  The resulting 5-quark ``tails" of the nucleon and Delta
baryons calculated in this way are given in the Appendix. The normalized distribution
of the 5-th body, namely $\bar u(x)$, over its momentum fraction is shown in Fig.\ref{fig_antiu}.
One can see a peak at $x_{\bar u}\sim 0.05$, which looks a generic phenomenon. The oscillations at large $x_{\bar u}$ reflect strong correlations in the wave function
between quarks, as well as perhaps indicate the insufficiently large functional basis used.
This part of the distribution is perhaps numerically unreliable.

%%%%%%%%%%%%%%%%%%%%%%%%%%%%%

\section{Perturbative and topology-induced antiquark sea}

%The  5-quark ($qqqq\bar q$) sector 
%we will focus on in this section is subleading in exclusive processes, but it is the first which
%introduce antiquarks. 
The results of our calculation cannot be directly compared to the   sea quark and antiquark PDFs, already plotted in Fig.\ref{fig_d_structure}, as those  include large perturbative contributions,
from gluon-induced quark pair production $g\rightarrow \bar u u, \bar d d $, dominant at very small $x$. However, these processes are basically flavor and chirality-independent,
while the observed flavor and spin asymmetries of the sea indicate that there must also
exist some nonperturbative mechanism of its formation. For a general recent review see
\cite{Geesaman:2018ixo}.

As originally emphasized by Dorokhov and Kochelev \cite{Dorokhov:1993fc},
The 't Hooft topology-induced 4-quark interaction leads to processes
$$ u \rightarrow u (\bar d d), \,\,\,\,  d \rightarrow d (\bar u u) $$ but not
$$u \rightarrow u (\bar u u),\,\,\,\,  d \rightarrow d (\bar d d) $$
which are forbidden by Pauli principle applied to zero modes. Since there are
two $u$ quarks and only one $d$ in the proton, one expects 
this mechanism to produce twice more $\bar d$ than $\bar u$.

\begin{figure}[htbp]
\begin{center}
\includegraphics[width=6cm]{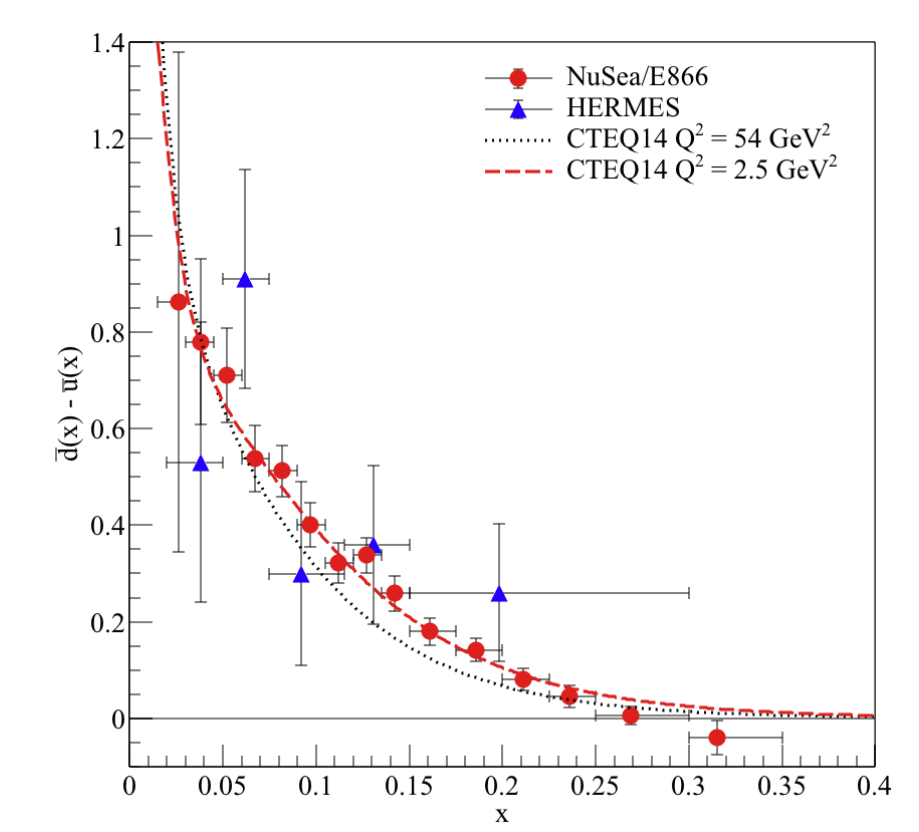}
\caption{The difference of sea antiquarks momentum distributions, $\bar d-\bar u$, from }
\label{fig_antid-antiu}
\end{center}
\end{figure}

The available experimental data, for the $difference$ of the sea antiquarks distributions  $\bar d-\bar u$ (from  \cite{Geesaman:2018ixo}) is shown in Fig.\ref{fig_antid-antiu}. In this difference
the symmetric gluon production should be cancelled out, and therefore it is
sensitive only to a non-perturbative contributions. 

Few comments: (i) First of all, the sign of the difference is indeed  as
predicted by the topological interaction, there are more anti-d than anti-u quarks;\\
(ii) Second, since 2-1=1, this representation of the data directly give us the nonperturbative antiquark production per valence quark, e.g. that of $\bar u$. This means it can be directly compared to the distribution we calculated from
the 5-quark tail of the nucleon and Delta baryons, Fig.\ref{fig_antiu}. \\
(iii)  The overall shape is qualitative similar, although our
calculation has a peak at $x_{\bar u}\sim 0.05$ while the experimental PDFs 
do not indicate it. Of course, there exist higher-quark-number sectors
with 7 and more quarks in baryons, which our calculation does not yet include: those should populate
the small $x$ end of the PDFs. \\
(iv) The data indicate much stronger decrease toward large   $x_{\bar u}$ than the calculation.\\
(v) There are other theoretical models which also reproduce the flavor asymmetry of the sea,
e.g. those with the pion cloud. In principle, one should be able to separate those and 
topology-induced mechanism (we focused to in this paper) by further combining flavor and
spin asymmetry of the sea. In particular, as also noticed in \cite{Dorokhov:1993fc}, if $d$ 
quark producing $u\bar u$ pair has positive helicity, the sea quark and antiquark from 't Hooft 4-quark operators must
have the $opposite$ (that is negative) helicity. The spin-zero pion mechanism,
on the other hand, cannot transfer spin and would produce flavor but not spin sea asymmetry.

\section{Summary and discussion} 

Application of some model Hamiltonians to light front wave functions of hadrons
should have been done long time ago. Yet it is just starting, with
the model presented in this paper still being (deliberately) rather schematic.

 We  
follow the approach of Ref.\cite{Jia:2018ary}, with the Hamiltonian
including three terms:
 (i) the ``constituent" light quark effective masses; (ii) the confinement term in  harmonic form; and (iii)
  local 4-quark NJL-type interaction, here reduced to its 't Hooft version. Its aim is to extend the model from meson to 
 baryon 3-quark
 states, then to 5-quark states, and finally to the mixing between the two sectors. 
 
  For this pilot study we make significant simplification of the model. We completely ignore transverse motion inside the hadrons, focusing on wave functions as a function of longitudinal momenta. We reduce the 
local 4-fermion NJL-type effective interaction to that  coming from 
gauge topology, believed to be the dominant part. 

 We have demonstrated that this simplified model does qualitatively describe the main features of
 meson and baryon mass splittings. The pion can become massless, while the $\eta'$ is heavier than the $\rho$ meson. The nucleon is lighter than $\Delta$ baryon, because of the attractive
spin-zero  $ud$ diquark channel. All of this is already known for quite some time, 
from Euclidean approaches such as instanton liquid model and lattice QCD.
 
 Our main finding  are however not the hadronic masses, but the light-front wave functions. 
We have shown that local 4-fermion residual interaction does indeed modify 
meson and baryon wave in a very substantial way. In \cite{Jia:2018ary}
  it was shown that the $\pi$ and $\rho$ meson have very different wave functions.
We now show that it is also true for
baryons: the proton and the $\Delta$ have qualitatively different wave functions.
 The imprint left by strong diquark ($ud$) correlations on the light-front wave functions
 is now established. We furthermore found evidences that Nucleon resonances 
 show features of ``quantum chaos" in quark motion. 
 
 We also made the first steps toward the ``un-quenching" the  light-front wave functions, estimating multiple
 matrix elements of the mixings between 3-q and 5-q (baryon-pentaquark) components of the nucleon wave function. 
This puts mysterious spin and flavor asymmetries of the nucleon sea inside the domain
of consistent Hamiltonian calculation.

Now, going to discussion part, let us point out several directions in which the model 
itself, as well as the calculations presented, 
can be improved.

As technical issues are concerned, 
the most direct (but important) generalization of this work would be to 
repeat these calculation in significantly larger functions space. 
In particular , one needs to restore dependence of the wave function on transverse motion, and on explicit spin states of the quarks. 

It would be just a straightforward exercise to generalize this work to strange mesons and baryons. We already discussed (around  Fig.\ref{fig_multiquark}) that $(\bar u u)(\bar d d)$
interaction includes comparable contributions from diagrams (b) and (c), while 
$(\bar s s)(\bar d d),(\bar u u)(\bar s s)$ only have the diagram (d): light quark masses
are too small to include sizable analogs of the diagram (b). Therefore $SU(3)_f$ 
breaking is expected to be substantial. 

As for the model, in its current form we only get a schematic understanding
of multi-quark correlations in the LFWFs.
One can use instead a more realistic NJL-type interactions, as used in modern PNJL
studies, or in works including its functional renormalization group flow, e.g. \cite{Drews:2016wpi}.
Yet, the main thing which needs improvement is to complement current
 local form of the 4-fermion interaction by a form with certain realistic formfactors.
 Even original NJL model had  a cutoff
parameter $\Lambda$, above which the interaction no longer exists. In the instanton liquid version, the formfactors have the form $\sim exp(-p \rho)$ where $\rho\sim 1/600\, MeV$
was the instanton size. The current model use purely local form, assuming
that topolgical solitons are small compared to hadronic sizes.Yet one needs these  formfactors, to be perhaps  able to finally find good 
convergence in the functional space of harmonics.

Needless to say, many applications of the LFWFs we did not even touched. For example,
one can calculate cross sections of various exclusive processes, e.g. electromagnetic formfactors or large angle scattering. Perhaps, one should only do so when the 
model used become more realistic and quantitative.

Finally, a comment about our general goal of bridging the  nonperturbative models
with observed PDFs and perturbative evolution. The pQCD
evolution of the baryon wave functions can be included if needed, as (matrix of) anomalous dimensions have been calculated quite some time ago \cite{Braun:1999te}. 
 The
multibody wave 
functions, if known, allows to do more than just predict the PDFs:   one may
in particular evaluate the matrix elements of the higher-twist operators.
So far, the only effort in this direction we are aware of, is Ref\cite{Braun:2011aw}
in which 4-body $q^3 g$ wave functions were derived and used for twist-3 estimates.
Comparing their properties with experiment at medium $Q^2$, would eventually provide 
a satisfactory ``bridge" between the perturbative and nonperturbative views on the
hadronic structure.

{\bf Acknowledgements.} The author thanks Felix Israilev for very useful discussion of
quantum chaos in manybody systems. The work is  supported  by the U.S. Department of Energy, Office of Science, under Contract No. DE-FG-88ER40388.

%%%%%%%%%%%%%%%%%%%%%

 \appendix
 \section{The 2-body LFWFs for pion, rho and eta' mesons}
 The approximate form of three mesonic wave functions are given below.
 The measure  times their squares is plotted in the Fig.\ref{fig_NJL_for_2}.
 The only comment is about convergence of the series: the first one, for $\rho$
 clearly shows sign of convergence, as the maximal coefficient is at the middle term.
 The pion seems to have a singularity at the ends points, while eta' has zero.
 
\be \psi_{\rho}(s) \approx -1.56263 + 13.5307 z^2 - 55.002 z^4 \ee
$$+ 138.857 z^6 - 241.184 z^8 + 
 298.688 z^{10} - 261.438 z^{12} + $$ $$153.563 z^{14} - 53.9465 z^{16} + 
 8.49524 z^{18}$$
 
 \be \psi_{\pi}(s) \approx -0.727959 - 3.13128 z^2 + 66.7135 z^4 \ee
$$  - 681.778 z^6 + 3689.13 z^8 - 
 11582.8 z^{10} + 21693.6 z^{12} $$ $$ - 23823.4 z^{14}  + 14124.3 z^{16} - 
 3485.37 z^{18} $$
  
\be   \psi_{\eta'}(s) \approx -1.77724 + 26.6155 z^2 - 147.594 z^4 \ee
$$+ 518.514 z^6 - 1365.06 z^8 + 
 2750.13 z^{10} - 3971.84 z^{12} $$ $$ + 3738.27 z^{14} - 2012.73 z^{16} + 
 465.809 z^{18} $$
 
 \section{The 3-body LFWFs for nucleon and Delta baryons}
 The 25 lowest harmonics shown in Fig.\ref{fig_harmonics} correspond to the following set of the $n,l$ values:
$$ (0, 0),
  (0, 1), (1, 1), (1, 0),
  (0, 2), (1, 2), (2, 2), (2, 0), $$
$$  (2, 1), (0, 3), (1, 3), (2, 3), (3, 3), (3, 2), (3, 1), (3, 0), $$
 $$ (0, 4), (1, 4), (2, 4), (3, 4), (4, 4), (4, 3), (4, 2), (4, 1), (4,   0)
$$
 The functional set is defined in  (\ref{eqn_conf_basis})
 
 Approximate polynomial form of the light front wave functions
 of the nucleon and Delta baryons, in $s,t$ variables
 $$ \psi_N\sim  -2.34 + 11.7 t + 43.5 t^2 - 4.78 t^3 - 48.7 t^4 + $$
\be  s^4 (-26.2 + 16.8 t + 316. t^2 + 14.2 t^3 - 
    523. t^4)  \ee $$+ 
 s (7.98 - 29.3 t + 31.4 t^2 + 115. t^3 - 
    97.3 t^4)$$ $$ + 
 s^3 (-12.4 + 57.4 t + 45. t^2 - 169. t^3 - 
    30. t^4) + $$ $$
 s^2 (22.3 - 32. t - 239. t^2 - 11.2 t^3 + 
    347. t^4) $$
% 
% 
% 
% 
% -4.94 + 8.66 t + 62.5 t^2 + 
%  1.87 t^3 - 72.5 t^4 + $$
% $$  s (3.63 - 31.1 t + 
%     61.7 t^2 + 115. t^3 - 
%     126. t^4) +  $$
% $$  s^2 (33.4 - 27.8 t - 
%     334. t^2 - 25.1 t^3 + 
%     460. t^4 $$
%   $$     s^3 (-8.51 + 58.4 t + 
%     19.3` t^2 - 162. t^3 - 
%     9.49 t^4) +  $$
%\be     +  s^4 (-36.4 + 16.4 t + 
%     411. t^2 + 26.6 t^3 - 
%     628. t^4) + 
% \ee
% 
 $$ \psi_\Delta\approx  -12.7 - 23.1 t + 39.9 t^2 + 36.2 t^3 - 
 39.5 t^4 +  $$
$$ + s^2 (39.2 + 69.9 t - 121. t^2 - 109. t^3 + 
    120. t^4) + $$
\be + s^4 (-28.1 - 49.4 t + 86.9 t^2 + 78.0 t^3 - 
    86.3 t^4) 
\ee
 \section{Models for nucleon LFWF used by Chernyak, Ogloblin and Zhitnitsky}
 In the paper \cite{Chernyak:1987nv} one finds usage at least three
 different nucleon LFWFs, which came form different usage of QCD sum rules method in 1980's.
 For definiteness, we reproduce it here
 $$
 N_{IJ} = 23.814 x_1^2 + 12.978 x_2^2 + 6.174 x_3^2 + 5.88 x_3 - 7.098$$
$$N_{KS} = 20.16 x_1^2 + 15.12 x_2^2 + 22.68 x_3^2 - 6.72 x_3 + 
   1.68 (x_1 - x_2) - 5.04;$$
\be N_J = 18.06 x_1^2 + 4.62 x_2^2 + 8.82 x_3^2 - 1.68x_3 - 2.94 \label{CS_wfs} \ee
 However, for purpose of comparison we used only one of them, $N_{IJ} $, in Fig.\ref{fig_Delta_N_CZ}, since
 they are all qualitatively similar.
 
  \section{The 5-body LFWF of the lowest pentaquark }
  For definiteness, we indicate the specific set of 5-body wave functions used.
  In the notations of (\ref{eqn_basis}) The set of states $(l,m,n,k)$ used in this section contains the following  states  for which $l+m+n+k<4$, 35 in total
$$
(0, 0, 0, 0),(1, 0, 0, 0),(0, 1, 0, 0), (0, 0, 1, 0), (0, 0, 0, 1), $$
$$  (1, 1, 0, 0), (1, 0, 1, 0), (1, 0, 0, 1), (0, 1, 1, 0), (0, 1, 0,    1),  $$
$$ (0, 0, 1, 1), (2, 0, 0, 0), (0, 2, 0, 0), (0, 0, 2, 0), (0, 0, 0, 2),
 $$ $$ (0, 1, 1, 1), (1, 0, 1, 1),   (1, 1, 0, 1), (1, 1, 1, 0), 
  (2, 1, 0, 0), $$ $$ (2, 0, 1, 0), (2, 0, 0, 1), (1, 2, 0, 0), 
  (0, 2, 1, 0), (0, 2, 0, 1),  $$ $$  (1, 0, 2, 0), (0, 1, 2, 0), (0, 0, 2, 1), 
 (1, 0, 0, 2), (0, 1, 0, 2),  $$ $$(0, 0, 1, 2),
  (3, 0, 0, 0), (0, 3, 0, 0), (0, 0, 3, 0), (0, 0, 0, 3)
$$
 The quark mass and confinement terms of the Hamiltonian,
  after diagonalization, produced the following
 mass spectrum (in $GeV^2$):
$$ M^2_{5q}\approx 13.9`, 13.7, 12.4, \
12.2, 11.0, 10.9, \
10.9, 10.8, 10.6, $$
$$10.6, 10.4, 10.3, \
10.3, 10.3, 10.3, \
10.1, 10.0, 9.97, $$
$$9.90, 9.76, 8.60, \
8.20, 8.14, 8.13, \
8.04, 8.02, 7.89, $$
\be 7.76, 7.71, 7.64, \
5.16, 5.05, 4.94, \
4.84, 4.04 \ee

The WF of them we will not present except of the lowest (the last). Its approximate form is
  $$ \psi_{penta}=1. - 1.21 t^3 + 1.61 u - 0.519 u^2 - 2.03 u^3 + $$
$$ s^2 (-1.79 + 0.0622 t + 0.177 u + 0.248 w) + $$
$$ t^2 (-1.40 + 0.265 u + 0.341 w) + 1.46 w - $$
$$ 0.488 u w + 0.510 u^2 w + 0.835 w^2 + $$
\be 0.194u w^2 - 2.81 w^3 + \ee
$$ t (1.19 - 0.117 u + 0.059 u^2 - 0.18 w - 
    0.018 u w + 0.074 w^2)
    $$
    where harmonics with smaller coefficients are neglected.
    
  \section{The LFWF in the 5-quark sector of the nucleon and Delta baryons} 
The 5-quark ``tails" of the nucleon and the Delta baryons we calculated has 
(in arbitrary normalization) the followiing wave functions 
 \be \psi_N^{5q}\approx -0.88 - 1.08 s^3 + 2.89 t^3 + \ee
$$ 3.55 u - 4.66 u^2 + 
  + $$
 $$t^2 (2.46  - 4.65 w) + $$
$$ s^2 (-0.44 - 0.42 t + 3.9 u - 1.83 w) - $$
$$ 0.40 w - 3.3 u w + 9.2 u^2 w + $$ $$4.45 w^2 - 
 7.4 u w^2 - 1. w^3 + $$
$$ t (0.94 - 4.32 u - 2.68 u^2 - 3.12 w + $$
  $$  10.1 u w ) + 
 s (1.86 + 1.96 t^2 - $$ $$1.914 u  + 
    t (0.25 - 2.48 u) $$ $$ - 7.0 w + 
    4.58 u w + 5.36 w^2) $$
    and 
     \be \psi_\Delta^{5q}\approx
     1.78 - 2.80 s^3 + 7.49 t^3 - \ee $$ 6.54 u + 3.33 u^2 + 
 s^2 (3.50 - 1.09 t $$ $$+ 0.91 u - 4.73 w) - 
 8.25 w + 20.1 u w  $$ $$- 1.71 u^2 w + 11.5 w^2 - 
 19.2 u w^2 -  $$ $$2.57 w^3 + t^2 (-5.35 + 1.19 w) + 
 t (-1.09 + 4.66 u -  $$ $$ 6.91 u^2 - 0.96 w + 
    1.44 u w) +  $$ $$
 s (0.05 - 0.20 t + 0.40 t^2 - 0.21 u +  $$ $$
    2.61 w + 0.35 u w - 1.08 w^2) $$

\end{document}